\documentclass[fleqn,12pt]{wlscirep}
\usepackage[utf8]{inputenc}
\usepackage[T1]{fontenc}
\usepackage{caption}
\usepackage{subcaption}
\usepackage{graphicx}
\usepackage{multibib}
\usepackage{soul}
\usepackage{color}
\usepackage{ulem}
\usepackage{threeparttable}
\usepackage{lineno}

\addto\captionsenglish{}

\newcommand{\Htwo}{H$_2$}

\newcommand{\Xco}{$X_{\rm CO}$}
\newcommand{\alphavir}{$\alpha_{\rm vir}$}
\newcommand{\thco}{$^{13}$CO}
\newcommand{\hii}{H\,{\sc ii}}
\newcommand{\Zsun}{Z$_{\odot}$}
\newcommand{\cmt}{cm$^{-3}$}

\newcommand{\nat}{Nature}
\newcommand{\apj}{Astrophys. J.}
\newcommand{\aap}{Astron. Astrophys.}
\newcommand{\aaps}{Astron. Astrophys. Suppl.}

\newcommand{\apjl}{Astrophys. J. Let.}
\newcommand{\apjs}{Astrophys. Js}
\newcommand{\araa}{Annu. Rev. Astron. Astrophys.}
\newcommand{\mnras}{Mon. Not. R. Astron. Soc.}
\newcommand{\pasp}{Publ. Astron. Soc. Pac.}

\title{Inadequate turbulent support in low-metallicity molecular clouds}

\author[1,2   ] {Lingrui Lin}
\author[1,2,* ] {Zhi-Yu Zhang}
\author[3,*   ] {Junzhi Wang}
\author[4,5   ] {Padelis P. Papadopoulos}
\author[1,2   ] {Yong Shi}
\author[6,7   ] {Yan Gong}
\author[7     ] {Yan Sun}
\author[1,2   ] {Yichen Sun}
\author[8     ] {Thomas G. Bisbas}
\author[9     ] {Donatella Romano}
\author[10,11,12 ] {Di Li}
\author[13,14 ] {Hauyu Baobab Liu}
\author[1,2   ] {Keping Qiu}
\author[15,16,17 ] {Lijie Liu}
\author[1,2,18] {Gan Luo}
\author[11,19,20 ] {Chao-Wei Tsai}
\author[11,20 ] {Jingwen Wu}
\author[21    ] {Siyi Feng}
\author[22    ] {Bo Zhang}

\affil[1]{School of Astronomy and Space Science, Nanjing University, Nanjing 210093, China}
\affil[2]{Key Laboratory of Modern Astronomy and Astrophysics (Nanjing University), Ministry of Education, Nanjing 210093, China}
\affil[3]{Guangxi Key Laboratory for Relativistic Astrophysics, School of Physical Science and Technology, Guangxi University, Nanning 530004, China}
\affil[4]{Department of Physics, Section of Astrophysics, Astronomy and Mechanics, Aristotle University of Thessaloniki, 54124 Thessaloniki, Greece}
\affil[5]{Research Center for Astronomy, Academy of Athens, Soranou Efesiou 4, GR-11527, Athens, Greece}
\affil[6]{Max-Planck-Institut f{\"u}r Radioastronomie, Auf dem H{\"u}gel 69, D-53121 Bonn, Germany}
\affil[7]{Purple Mountain Observatory and Key Laboratory of Radio Astronomy, Chinese Academy of Sciences, Nanjing 210008, China}
\affil[8]{Research Center for Astronomical Computing, Zhejiang Laboratory, Hangzhou 311100, China}
\affil[9]{INAF, Astrophysics and Space Science Observatory, Via Gobetti 93/3, I-40129 Bologna, Italy}
\affil[10]{Department of Astronomy, Tsinghua University, Beijing 100084, China}
\affil[11]{National Astronomical Observatories, Chinese Academy of Sciences, A20 Datun Road, Chaoyang District, Beijing 100101, China}
\affil[12]{Zhejiang Lab, Hangzhou, Zhejiang 311121, People’s Republic of China}
\affil[13]{Department of Physics, National Sun Yat-Sen University, No. 70, Lien-Hai Road, Kaohsiung City 80424, Taiwan}
\affil[14]{Center of Astronomy and Gravitation, National Taiwan Normal University, Taipei 116, Taiwan}
\affil[15]{Cosmic Dawn Center (DAWN), Copenhagen, Denmark}
\affil[16]{Niels Bohr Institute, University of Copenhagen, Lyngbyvej 2, 2100 Copenhagen \O, Denmark}
\affil[17]{DTU-Space, Technical University of Denmark, Elektrovej 327, DK2800 Kgs. Lyngby, Denmark}
\affil[18]{Institut de Radioastronomie Millimetrique, 300 rue de la Piscine, Domaine Universitaire de Grenoble, F-38406 Saint-Martin d'H{\'e}res, France}
\affil[19]{Institute for Frontiers in Astronomy and Astrophysics, Beijing Normal University, Beijing 102206, China}
\affil[20]{School of Astronomy and Space Science, University of Chinese Academy of Sciences, Beijing 100049, China}
\affil[21]{Department of Astronomy, Xiamen University, Zengcuo’an West Road, Xiamen, 361005, China}
\affil[22]{Shanghai Astronomical Observatory, Chinese Academy of Sciences, 80 Nandan Road, Shanghai 200030, China}

\affil[*]{corresponding author, zzhang@nju.edu.cn, junzhiwang@gxu.edu.cn}

\clearpage

\begin{abstract}
The dynamic properties of molecular clouds are set by the interplay of their self-gravity, turbulence, external pressure and magnetic fields. Extended surveys of Galactic molecular clouds typically find that their kinetic energy ($E_{\rm k}$) counterbalances their self-gravitational energy ($E_{\rm g}$), setting their virial parameter $\alpha_{\rm vir}=2E_{\rm k}/|E_{\rm g}|\approx1$. However, past studies either have been biased by the use of optically-thick lines or have been limited within the solar neighborhood and the inner Galaxy (Galactocentric radius $R_{\rm gc}<R_{\rm gc,\odot} \approx 8$ kpc). Here we present sensitive mapping observations of optically thin \thco\ lines towards molecular clouds in the low-metallicity Galactic outer disk ($R_{\rm gc}\sim9-24$ kpc). By combining archival data from the inner Galaxy and four nearby metal-poor dwarf galaxies, we reveal a systematic trend of $\alpha_{\rm vir}$, which declines from supervirial dynamic states in metal-rich clouds to extremely subvirial dynamic states in metal-poor clouds. In these metal-poor environments, turbulence alone is insufficient to counterbalance the self-gravity of a cloud. A cloud-volumetric magnetic field may replace turbulence as the dominant cloud-supporting mechanism in low-metallicity conditions, for example, the outermost galactic disks, dwarf galaxies and galaxies in the early Universe, which would then inevitably impact the initial conditions for star formation in such environments.
\end{abstract}

\begin{document}

\maketitle

\thispagestyle{empty}

Larson's relations\cite{1981MNRAS.194..809L} have been well established both in the Milky Way \cite{1981MNRAS.194..809L,1987ApJ...319..730S,2009ApJ...699.1092H,2017ApJ...834...57M} and in external galaxies \cite{2008ApJ...686..948B,2010ARA&A..48..547F,2010MNRAS.406.2065H,2017ApJ...835..278S,2019ApJ...885...50W,2023A&A...672A.153S}. They consist of power-law relations among the size, velocity dispersion, and mass of molecular clouds.  Such relations are often considered as the outcome of virial equilibrium \cite{1981MNRAS.194..809L, 1989ApJ...338..178E} (or energy equipartition \cite{2006MNRAS.372..443B}; Methods) between the turbulent kinetic energy ($E_{\rm k}$) and the self-gravitational energy ($E_{\rm g}$) of a molecular cloud, which is characterized by the virial parameter ($\alpha_{\rm vir} = 2E_{\rm k}/|E_{\rm g}|$)\cite{1992ApJ...395..140B} that is typically near unity\cite{1981MNRAS.194..809L,1987ApJ...319..730S,2009ApJ...699.1092H}. Molecular clouds resolved in the Galactic Centre and in some external galaxies seem to have a slight preference towards supervirial states ($\alpha_{\rm vir}>$1), possibly due to external pressure on cloud boundaries \cite{2018ApJ...860..172S} or tidal shear around the cloud envelopes \cite{2016ApJ...832..143F,2021MNRAS.505.4048L,2023MNRAS.525..962P}. The virial equilibrium assumption has been used in a variety of contexts, such as for calibrating the standard value of the so-called CO-to-H$_{2}$ conversion factor ($X_{\rm CO}$) \cite{2013ARA&A..51..207B}, for estimating the average turbulent gas pressure in high-redshift galaxies \cite{2011ApJ...742...11S} and for setting the initial conditions of star formation in numerical simulations \cite{2003ApJ...585..850M,2017A&A...605A..97V}.

Nevertheless, most previous studies on the dynamic states of molecular clouds were based solely on the more luminous but optically thick low-$J$ $^{12}$CO transitions \cite{1987ApJ...319..730S,2017ApJ...834...57M}. The uncertainty of $X_{\rm CO}$ values \cite{2013ARA&A..51..207B}, the opacity broadening \cite{2016A&A...591A.104H} and the radiative trapping of $^{12}$CO lines (Methods; the latter can keep $^{12}$CO lines luminous even for low-density outer-cloud envelope gas, which may not be gravitationally bound) undoubtedly complicate $\alpha_{\rm vir}$ estimates for molecular clouds. 

Rotational transitions of the rarer $^{13}$CO molecule \cite{1994ARA&A..32..191W}, on the other hand, are optically thin for the bulk of the mass in molecular clouds in most cases \cite{2009ApJS..182..131R}. Therefore, the low-$J$ $^{13}$CO lines can more faithfully trace \Htwo\ mass and velocity dispersion \cite{2009ApJ...699.1092H} than the $^{12}$CO lines. However, because of their faintness, studies of cloud dynamics based on $^{13}$CO lines are mostly limited to within a Galactocentric radius ($R_{\rm gc}$) of less than 12 kpc (refs.~\citen{2001ApJ...551..852H,2021A&A...654A.144B}), which includes the solar neighbourhood \cite{1981MNRAS.194..809L} and the inner Galaxy \cite{2009ApJ...699.1092H,2010ApJ...723..492R,2019A&A...632A..58R} but leaves the Galactic outer disk less explored.

The Galactic outer disk, namely, the portion of the disk beyond the solar circle ($R_{\rm gc}>R_{\rm gc,\odot} \approx 8$ kpc)\cite{knapen2017outskirts}, features very different physical conditions, such as low midplane pressure\cite{2003ApJ...587..278W}, a low turbulence injection level \cite{2017ApJ...834...57M} and low gas-phase metallicity ($Z$, routinely traced by the oxygen abundance, O/H)\cite{2022MNRAS.510.4436M}. Such environments reflect not only the early formation stage of the Galactic thin disk\cite{chiappini2008chemical} but also the general properties of gas-rich, metal-poor galaxies \cite{2014Natur.514..335S,2014A&A...561A..49H}. 

\begin{figure}[t!]
\includegraphics[width=1.0\linewidth]{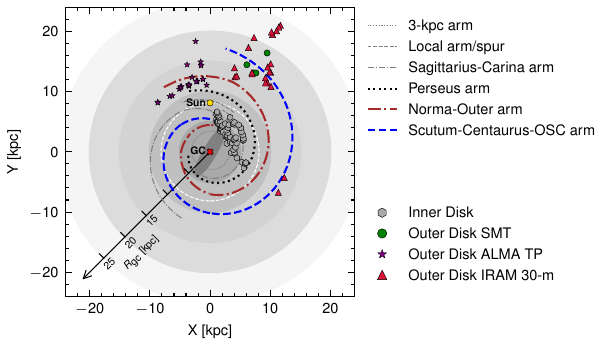}
	\caption{\textbf{Distribution of molecular clouds on the Galactic plane.} Spiral arm models from the Bar and Spiral Structure Legacy (BeSSeL) Survey\cite{2019ApJ...885..131R} are overlaid, including the 3-kpc arm (gray dotted line), the Local arm/spur (gray dashed line), the Sagittarius–Carina arm (gray dash-dotted line), the Perseus arm (black thick dotted line), the Norma-Outer arm (brown thick dash-dotted line), and the Scutum–Centaurus–OSC arm (blue thick dashed line). The gold dot and the red cross locate the Sun and the Galactic center (GC), respectively. The white dashed line shows the Solar circle ($R_{\rm gc,\odot}$). The gray-shadowed regions represent Galactocentric radii of 5, 10, 15, 20, and 25 kpc. Purple stars, red triangles, and green dots are the outer disk molecular clouds observed with ALMA Total Power, IRAM 30-m, and SMT, respectively. The gray hexagons show inner disk molecular clouds from the Galactic Ring Survey \cite{2009ApJ...699.1092H}.}
\label{fig: Main_Fig1}
\end{figure}

In this work, we performed mapping observations of the $^{13}$CO $J=1\rightarrow0$ and $J=2\rightarrow1$ transitions for a sample of molecular clouds in the Galactic outer disk ($9<R_{\rm gc}<25$ kpc), using the Institut de radioastronomie millimétrique (IRAM) 30 m telescope, the single-dish telescope (Total Power (TP) Array) of the Atacama Large Millimeter/submillimeter Array (ALMA) and the Submillimeter Telescope (SMT). Figure~\ref{fig: Main_Fig1} shows the spatial distribution of our samples on the Galactic plane.

We first derived the surface density of $^{13}$CO ($N_{\rm ^{13}CO}$) from the $^{13}$CO emission and then obtained the \Htwo\ surface density ($N_{\rm H_{2}}$) with the \Htwo/\thco\ abundance ratio estimated from the Galactic radial gradients of $^{12}$C/$^{13}$C (ref.~\citen{2020A&A...640A.125J}) and O/H (ref.~\citen{2022MNRAS.510.4436M}; Methods). For each cloud, we measured the equivalent cloud radius ($R_{\rm cloud}$), the velocity dispersion ($\sigma_{\rm v}$) and the molecular gas mass ($M_{\rm mol}$), all within the half-peak isophote of $N_{\rm H_{2}}$ (Methods). Cloud distances were estimated based on the Galactic rotation curve model from the Bar and Spiral Structure Legacy (BeSSeL) survey \cite{2019ApJ...885..131R}.

For comparison, we retrieved archival $^{13}$CO $J=1\rightarrow0$ and $J=2\rightarrow1$ data from the inner Galaxy \cite{2009ApJ...699.1092H} and nearby metal-poor dwarf galaxies, namely, the Large Magellanic Cloud \cite{2019ApJ...885...50W} (LMC; $Z\approx0.5$ \Zsun; ref.~\citen{1992ApJ...384..508R}), the Small Magellanic Cloud \cite{2018ApJ...853..111J} (SMC; $Z\approx0.2$ \Zsun; ref.~\citen{2003ASPC..304..187P}) and NGC~6822 ($Z\approx0.2$ \Zsun) \cite{2016A&A...586A..59G}. We also took the physical properties of molecular clumps from an extremely metal-poor dwarf galaxy DDO~70 ($Z\approx0.07$ \Zsun) from the literature\cite{2020ApJ...892..147S}.

\begin{figure}[t!]
 \includegraphics[width=0.5\textwidth]{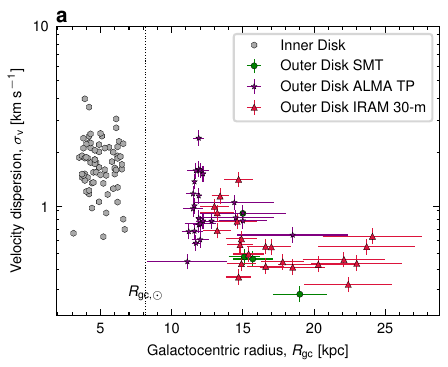}
 \includegraphics[width=0.5\textwidth]{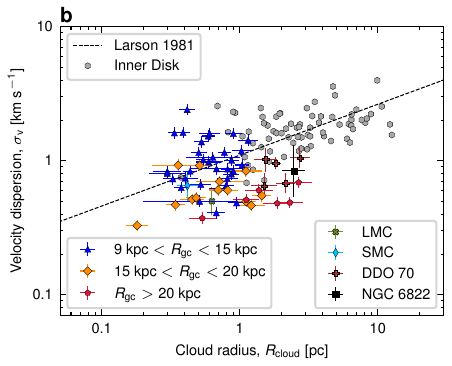}
\caption{\textbf{Variations of the cloud velocity dispersion.} \textbf{a,} $\sigma_{\rm v}$ versus $R_{\rm gc}$. Purple stars, red triangles, and green dots are the outer disk molecular clouds observed with ALMA Total Power, IRAM 30-m, and SMT, respectively. The gray hexagons show inner disk molecular clouds from the Galactic Ring Survey \cite{2009ApJ...699.1092H}. The vertical black-dotted line shows the Galactocentric radius of the Sun ($R_{\rm gc,\odot}$). \textbf{b,} $\sigma_{\rm v}$ versus $R_{\rm cloud}$. The outer disk molecular clouds are divided into three $R_{\rm gc}$ bins: 9 kpc $<R_{\rm gc}<$ 15 kpc (blue triangles), 15 kpc $<R_{\rm gc}<$ 20 kpc (orange diamonds), and $R_{\rm gc}>$ 20 kpc (red pentagons). Molecular clouds from metal-poor galaxies, i.e, LMC (olive cross, median), SMC (blue thin-diamond, median), NGC~6822 (black squares, median) and DDO 70 (brown pluses), are overlaid. The dashed line shows the classical Larson's $\sigma_{\rm v}-R_{\rm cloud}$ relation \cite{1981MNRAS.194..809L}. Data are presented as measured values with 1-$\sigma$ uncertainties.}
\label{fig: Main_Fig2}
\end{figure}

Figure~\ref{fig: Main_Fig2}a shows that the velocity dispersion $\sigma_{\rm v}$ of the Galactic molecular clouds decreases from $R_{\rm gc}=5$ kpc to 15 kpc, with this trend becoming flat at $R_{\rm gc}>15$ kpc. This is consistent with the results from a low-resolution $^{12}$CO $J=1\approx0$ survey on a tens of parsecs scale \cite{2017ApJ...834...57M} and may be related to the decreasing kinetic energy injection towards the outer Galaxy (Methods). Figure~\ref{fig: Main_Fig2}b presents $\sigma_{\rm v}$ versus $R_{\rm cloud}$, showing that, although molecular clouds at $R_{\rm gc}<$ 15 kpc roughly follow the classical Larson's $\sigma_{\rm v}$ versus $R_{\rm cloud}$ relation \cite{1981MNRAS.194..809L}, those at $R_{\rm gc}>$ 15 kpc and those from metal-poor dwarf galaxies have a lower $\sigma_{\rm v}$ than what would be expected from Larson's relation. Such a discrepancy has also been revealed by $^{12}$CO line observations of metal-poor dwarf galaxies, including the Magellanic Clouds \cite{2008ApJ...686..948B, 2023ApJ...949...63O}.

\begin{figure}[t!]
 \includegraphics[width=1.0\textwidth]{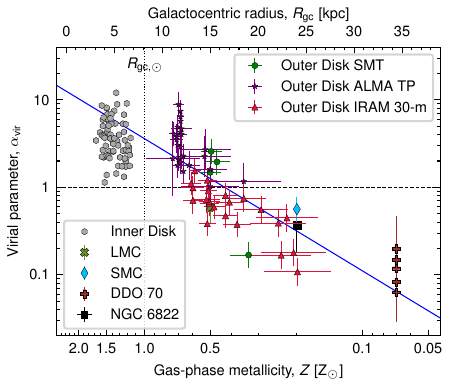}
\caption{\textbf{Variation of cloud virial parameter as a function of the gas-phase metallicity.} Purple stars, red triangles, and green dots are the outer disk molecular clouds observed with ALMA Total Power, IRAM 30-m, and SMT, respectively. The gray hexagons show inner disk molecular clouds from the Galactic Ring Survey \cite{2009ApJ...699.1092H}. Molecular clouds from metal-poor galaxies, i.e, LMC (olive cross, median), SMC (blue thin-diamond, median), NGC~6822 (black squares, median) and DDO 70 (brown pluses), are overlaid. The black-dashed line represents the virial equilibrium between the kinetic energy and the self-gravitational energy, namely $\alpha_{\rm vir}=1$. The vertical black dotted line shows the Galactocentric radius of the Sun ($R_{\rm gc,\odot}$). For molecular clouds in the Milky Way, the gas-phase metallicity ($Z$, bottom axis) is estimated from the Galactocentric radius ($R_{\rm gc}$, top axis) through the O/H-$R_{\rm gc}$ gradient \cite{2022MNRAS.510.4436M}. The blue line shows a linear fitting of the \alphavir$-Z$ trend using the \thco\ data in the Milky Way (see Methods). Data are presented as measured values with 1-$\sigma$ uncertainties.}
\label{fig: Main_Fig3}
\end{figure}

Figure~\ref{fig: Main_Fig3} shows that the $\alpha_{\rm vir}$ of molecular clouds, from both the Milky Way and metal-poor dwarf galaxies, systematically varies as a function of the gas-phase metallicity. The gas-phase metallicities of the Galactic clouds were estimated using the Galactic radial gradient of O/H (ref.~\citen{2022MNRAS.510.4436M}). $\alpha_{\rm vir}$ decreases from supervirial states ($\alpha_{\rm vir}>1$) in the metal-rich inner Galaxy to subvirial states ($\alpha_{\rm vir}<1$) towards the metal-poor outer Galaxy with $R_{\rm gc}>15$ kpc ($Z<0.5$ \Zsun). The slope of this correlation is sensitive to the adopted \Htwo/\thco\ abundance ratio gradient. Nevertheless, the overall trend is robust due to the monotonically decreasing Galactic radial gradients for both $^{12}$C/$^{13}$C and the gas-phase metallicity. Molecular clouds from metal-poor dwarf galaxies, that is, the LMC ($Z\approx0.5$ \Zsun)\cite{1992ApJ...384..508R}, the SMC ($Z\approx0.2$ \Zsun)\cite{2003ASPC..304..187P}, NGC~6822 ($Z \approx 0.2$ \Zsun)\cite{2016A&A...586A..59G} and DDO~70 ($Z \sim 0.07$ \Zsun)\cite{2020ApJ...892..147S}, extend the \alphavir versus $Z$ trend to the extremely low-metallicity end, despite that these systems have a variety of stellar contents, star-formation properties and galactic dynamic conditions. Consequently, we found that the classical virial equilibrium between $E_{\rm k}$ and $E_{\rm g}$ of molecular clouds is not universal.

Although by definition \alphavir\ includes only $E_{\rm k}$ and $E_{\rm g}$, the general virial theorem \cite{spitzer2008physical,1989ApJ...338..178E,1992ApJ...395..140B} (Methods) also contains terms dependent on the volumetric magnetic field ($B$) and the cloud-boundary pressure ($P_{\rm e}$). These other mechanisms can support and confine molecular clouds, respectively \cite{1989ApJ...338..178E,1992ApJ...395..140B}. 

The subvirial clouds found at $Z<0.5$ \Zsun\ ($R_{\rm gc}>15$ kpc) indicate that the turbulent velocity fields alone are not sufficient to counterbalance the cloud self-gravity in such low-metallicity conditions. To sustain the survival of the clouds, another mechanism is necessary. The most plausible mechanism appears to be the magnetic field (Methods).

Using the general virial theorem and assuming efficient momentum transfer between ions and neutral species \cite{1998ApJ...503..689W}, we found that the $B$-field strengths required to support subvirial clouds are consistent with expectations. These are based on the $B$ versus $n_{\rm H}$ relation benchmarked in the solar neighborhood and the inner Galaxy \cite{2010ApJ...725..466C} ($n_{\rm H}$, volume density of hydrogen nuclei; Methods). Therefore, the subvirial clouds in metal-poor conditions do not require a stronger $B$ field than those in the metal-rich inner Galaxy. At low metallicities, the same $B$-field strength can overtake turbulence in balancing the self-gravity of a cloud. 

The conditions necessary for such subvirial states may arise from the distinct physical properties typical of low-metallicity environments (Methods). These include reduced turbulence in atomic gas \cite{2023ApJ...954...38K} and a higher ionization fraction in molecular gas. The latter would enhance the coupling between $B$-field lines and gas\cite{lequeux2004interstellar}, allowing the magnetic field to play a more dominant role in counteracting self-gravity compared to high-metallicity regimes.

The supervirial molecular clouds from the metal-rich inner Galaxy, on the other hand, are probably the result of enhanced gas pressures at cloud boundaries ($P_{\rm e}/k\approx10^{4}-10^{7}$\,K\,cm$^{-3}$), as previously proposed in the literature \cite{2011MNRAS.416..710F}. Other factors, such as cloud-boundary magnetic fields, tidal effects \cite{2016ApJ...832..143F,2021MNRAS.505.4048L,2023MNRAS.525..962P} and even embedded stellar objects \cite{2022MNRAS.515.2822R}, can all drive \alphavir\ above unity (Methods).

Regardless of the detailed mechanisms driving the supervirial and subvirial conditions, the dynamic states of molecular clouds seem to depend strongly on the ambient environments. This will inevitably impact all star formation theories, given that the dynamic states of molecular clouds are crucial aspects of the initial conditions for star formation in galaxies. The subvirial states found in low-$Z$ molecular clouds, in particular, are expected to be consequential for star formation in several environments, such as the outskirts of main-sequence galaxies (still metal-poor), dwarf galaxies and galaxies at cosmic dawn \cite{2023NatAs...7..622C}.

\setcounter{figure}{0}
\captionsetup[figure]{labelformat=simple, labelsep=colon, name=Supplementary Fig.}
\setcounter{table}{0}
\captionsetup[table]{labelformat=simple, labelsep=colon, name=Supplementary Table}

\section*{Methods}

\subsection*{Observations towards the Galactic outer disk clouds}

\noindent \textbf{$\bullet$ IRAM 30 m telescope observations.} Observations with the IRAM 30 m telescope were executed from 29 August to 5 September 2017 (Project ID 031-17, principal investigator (PI) Z.-Y. Zhang) and from 3 to 30 Janunary 2023 (Project ID 102-22, PI T. Bisbas). The targets were selected from the literature \cite{1989A&AS...80..149W,2015ApJ...798L..27S,2016RAA....16...47L,2017ApJS..230...17S}, based on both their $^{12}$CO brightness and Galactocentric radius.

In Project 031-17, we performed on-the-fly (OTF) mapping observations towards 14 molecular clouds in the outer Galactic disk. We used the Eight MIxer Receiver (EMIR) at 3 mm (E0), which was equipped with the fast Fourier transform spectrometer in FTS200 mode, to target the $^{13}$CO $J=1\rightarrow0$ transition. The spectral resolution was 195 kHz, corresponding to a velocity resolution of $\sim 0.53 \; \rm km\;s^{-1}$ at the rest frequency of $^{13}$CO $J=1\rightarrow0$ (110.20135430 GHz). The mapping area towards each target was $2.5'\times2.5'$ along both right ascension and declination. The step between each OTF scan of rows and columns was $9''$, and the separation between each integration was $4.8''$, enabling a super-Nyquist sampling of the half power beam-width ($\theta_{\rm beam}\approx23.5''$). 

The OTF mapping data were reduced with \texttt{GILDAS/CLASS}. Platforming effects were firstly removed by separately fitting and subtracting the baselines of each FTS200 unit. We extracted a velocity interval centred at the $^{13}$CO $J=1\rightarrow0$ line with a width of $\sim18$ times the full width of half maximum linewidth. We used the \texttt{xy\_map} task to build regular $^{13}$CO spectral cubes with pixel sizes of 4$''\times$4$''$. Given a main beam efficiency ($\eta_{\rm beam}$) of 0.78 (\url{https://publicwiki.iram.es/Iram30mEfficiencies}), the typical root-mean-square (r.m.s.) main beam temperature ($T_{\rm mb,rms}$) was $\sim$ 0.13 K.

In Project 102-22, we mapped ten outer disk molecular clouds, covering the $J=1\rightarrow0$ lines of $^{12}$CO and $^{13}$CO using the E0 receiver. The step between each OTF scan of rows and columns was $4.5''$ and the separation between each integration was $2.6''$. We reduced the data and built the spectral cubes following the same procedure used for Project 031-17. The typical $T_{\rm mb,rms}\approx0.08$ K. Two of the ten targets were excluded in this work due to their low signal-to-noise ratio ($S/N$).

\noindent \textbf{$\bullet$ ALMA TP Array observations.} The ALMA observations (Project ID 2021.2.00175.S, PI L. Lin) were executed during ALMA cycle 8 from 29 January to 28 September 2022. The $J=2\rightarrow1$ transitions of $^{12}$CO and $^{13}$CO were covered. The targets were selected from the literature \cite{1989A&AS...80..149W}. We obtained $4'\times4'$ OTF maps towards 26 outer-disk molecular clouds in the third Galactic quadrant ($180^{\circ}<l<270^{\circ}$ where $l$ is the Galactic longitude) with the ALMA TP Array. The data were reduced with the standard pipeline. The $\theta_{\rm beam}$ and the velocity resolution at the rest frequency of $^{13}$CO $J=2\rightarrow1$ (220.39868420 GHz) were $\sim29.5''$ and 0.16 $\rm \;km\;s^{-1}$, respectively. The typical $T_{\rm mb,rms}\approx0.04$ K.

\noindent \textbf{$\bullet$ SMT observations.} The SMT observations (Project ID Lin\_L\_22B\_1, PI L. Lin) were executed from 29 October to 1 November 2022. We obtained $4'\times4'$ OTF maps towards 13 outer-disk molecular clouds\cite{1989A&AS...80..149W,2015ApJ...798L..27S}, nine of which had $^{13}$CO $J=1\rightarrow0$ observations made by IRAM 30 m telescope. Therefore, only four targets are included in this work. The $J=2\rightarrow1$ transitions of $^{12}$CO and $^{13}$CO were covered. The step between each OTF scan of rows/columns was $10''$, and the separation between each integration was $\sim1.3''$, enabling a super-Nyquist sampling of $\theta_{\rm beam}\approx36''$. The velocity resolution was 0.34 $\rm \;km\;s^{-1}$. Given $\eta_{\rm beam}=0.70$, the typical $T_{\rm mb,rms}\approx0.14$ K. 

The numbers of molecular clouds observed by us at 9 kpc $\leq R_{\rm gc}<$ 15 kpc, 15 kpc $\leq R_{\rm gc}<$ 20 kpc, and $R_{\rm gc}\geq$ 20 kpc are 33, 13, and six, respectively.

\subsection*{Archival data}

\noindent \textbf{$\bullet$ Galactic Ring Survey $^{13}$CO $J=1\rightarrow0$ data from the inner Galaxy.} We retrieved the $^{13}$CO $J=1\rightarrow0$ data for a sample of inner disk molecular clouds \cite{2009ApJ...699.1092H} studied by the Galactic Ring Survey (GRS)\cite{2006ApJS..163..145J}. The spatial sampling step of GRS was $22''$, enabling a Nyquist sampling of the telescope beam ($\theta_{\rm beam}=46''$)\cite{2006ApJS..163..145J}. Given a main beam efficiency $\eta_{\rm mb}=0.46$, the typical r.m.s. main beam temperature is 0.27 K at the velocity resolution of 0.21 km s$^{-1}$. We visually excluded clouds suffering from severe line-of-sight (LoS) cloud blending. Our measurements are consistent with previous work \cite{2009ApJ...699.1092H}.

\noindent \textbf{$\bullet$ ALMA archival data of nearby galaxies.} With the capability of the ALMA 12-m array, molecular clouds in the nearby galaxies can be resolved on parsec scales. 

\begin{itemize}

\item[$\triangleright$] {\bf LMC.} We used the $^{13}$CO cubes provided in the literature\cite{2019ApJ...885...50W} for several regions in the LMC. Specifically, we included the $^{13}$CO $J=1\rightarrow0$ cubes (ALMA 12 m only) for N59C, A439, GMC104 and GMC1, and the $^{13}$CO $J=2\rightarrow1$ cube (ALMA 12 m plus the TP Array) for the Planck cold cloud. All cubes have a restored $\theta_{\rm beam}=3.5''$, corresponding to a physical resolution of 0.8 pc at the distance of the LMC. The r.m.s noise of the brightness temperature was $\sim0.2$ K at a velocity resolution of 0.2 km s$^{-1}$.

\item[$\triangleright$] {\bf SMC.} We retrieved ALMA $^{13}$CO $J=2\rightarrow1$ data from the ALMA archival system (Project IDs 2013.1.00652.S and 2015.1.00581.S)\cite{2018ApJ...853..111J}. Four regions in the SMC (N22, SWBarN, SWBarS and SWDarkPK) were observed using the Atacama Compact Array (7 m array plus the TP Array; 2013.1.00652.S) and the ALMA 12 m array (2015.1.00581.S). The interferometry data (12 m plus 7 m arrays) were calibrated with the \textsc{CASA} pipeline (v4.5.2-r36115 and v4.2.2-r30986). The imaging was done by \texttt{tclean} in \textsc{CASA} v6.5.3. The ideal angular resolution of these observations was $\sim1''$. For consistency, however, we imaged the interferometry data with $weighting=`briggs$' ($robust=2.0$), $cell=0.4''$ and $restoringbeam=2.0''$. The channel width of the cleaned cubes was 0.2 km s$^{-1}$. The interferometry cubes were combined with the product TP cubes using the \texttt{feather} task. The typical r.m.s. brightness temperature was $\sim0.4$ K ($\sim$ 0.06 Jy per beam).

\item[$\triangleright$] {\bf NGC~6822.} For NGC~6822, we retrieved the $^{13}$CO $J=1\rightarrow0$ data observed by ALMA 12 m array (Project ID 2019.1.01641.S). We calibrated the data with the standard \textsc{CASA} pipeline. The imaging was done by \texttt{tclean} in \textsc{CASA} v6.5.3. For a compromise between angular resolution and sensitivity, we imaged the data with $weighting=`briggs$' ($robust=2.0$), $cell=0.3''$ and $restoringbeam=2.4''$. The channel width was $\sim$ 2.7 km s$^{-1}$. The typical r.m.s.  brightness temperature was $\sim0.01$ K ($\sim$ 0.6 mJy per beam).

We visually identified molecular clouds in the LMC, SMC and NGC~6822. To exclude clouds that were mixed with several components along the LoS, we kept only those isolated clouds for which the half-peak isophote of $N_{\rm H_{2}}$ was a single closed contour. Clouds with radii smaller than half of the resolution beam were excluded. 

\item[$\triangleright$] {\bf DDO~70.} We included five molecular clumps detected in a local extremely metal-poor galaxy, DDO~70 ($Z\approx0.07$ \Zsun, at a distance of 1.38 Mpc) \cite{2020ApJ...892..147S}. Even though these clouds were identified through the $^{12}$CO $J=2\rightarrow1$ emission, the cloud properties were measured independently of the uncertain $X_{\rm CO}$ (ref.~\cite{2020ApJ...892..147S}).

\end{itemize}

Supplementary Table~\ref{tab: ED_Table1} summarizes the details of the above new observations and archival data.

\begin{table*}[h!]
\centering
\caption{{\bf Information about the cloud samples.} $R_{\rm gc}$: Galactocentric radius; $Z$: gas-phase metallicity; $\Delta V_{\rm ins}$: instrumental broadening, i.e., channel width; $\theta_{\rm beam}$: half-power beam width; $T_{\rm b,rms}$: root-mean-square noise of main beam temperature; Phy. Res.: physical resolution.}
\label{tab: ED_Table1}
\begin{threeparttable}
\begin{tabular}{ccccccccc}
\hline
Location                  & $R_{\rm gc}$ & $Z$       & $^{13}$CO transition          & Number & $\Delta V_{\rm ins}$ & $\theta_{\rm beam}$ & $T_{\rm b,rms}$ & Phy. Res.     \\
                          & (kpc)        & (\Zsun)   & ($J\rightarrow\,J-1$)         &        & ($\rm \;km\;s^{-1}$) & ($''$)              & (K)             & (pc)     \\
\hline
Outer Galaxy\tnote{a}     & 12.2 - 24.0  & 0.2 - 0.7 & $1\rightarrow0$               & 14     & 0.53                 & 23.5                & 0.13            & 0.56 - 2.0     \\
Outer Galaxy\tnote{b}     & 16.1 - 23.0  & 0.2 - 0.4 & $1\rightarrow0$               & 8      & 0.53                 & 23.5                & 0.08            & 1.1 - 1.9     \\
Outer Galaxy\tnote{c}     & 9.8 - 17.8   & 0.4 - 0.8 & $2\rightarrow1$               & 26     & 0.16                 & 29.5                & 0.04            & 0.24 - 1.4     \\
Outer Galaxy\tnote{d}     & 15.7 - 16.6  & 0.4 - 0.5 & $2\rightarrow1$               & 4      & 0.34                 & 36.0                & 0.14            & 1.7 - 1.8     \\
Inner Galaxy\tnote{e}     & 3.2 - 7.4    & 1.1 - 1.7 & $1\rightarrow0$               & 72     & 0.21                 & 46.0                & 0.27            & 0.56 - 2.0    \\
LMC\tnote{f}              & ---          & 0.5       & $1\rightarrow0/2\rightarrow1$ & 62     & 0.20/0.50            & 3.5                 & 0.2             & 0.85   \\
SMC\tnote{g}              & ---          & 0.2       & $2\rightarrow1$               & 37     & 0.20                 & 2.0                 & 0.6             & 0.60 \\
NGC~6822\tnote{h}         & ---          & 0.2       & $1\rightarrow0$               & 6      & 2.7                  & 2.4                 & 0.01            & 5.5 \\
DDO~70\tnote{i}           & ---          & 0.07      & ---                           & 5      & 0.4                  & 0.2                 & 0.5             & 1.47 \\
\hline
\end{tabular}
\footnotesize{$^a$ IRAM 30-m (2017),  $^b$ IRAM 30-m (2023),  $^c$ ALMA Total Power (Cycle 8),  $^d$ SMT (2022),  $^e$ FCRAO 14-m (The Galactic Ring Survey)\cite{2006ApJS..163..145J,2009ApJ...699.1092H},  $^f$ ALMA 12-m/7-m/Total Power LMC\cite{2019ApJ...885...50W},  $^g$ ALMA 12-m/7-m/Total Power SMC\cite{2018ApJ...853..111J},  $^h$ ALMA 12-m,  $^i$ ALMA 12-m\cite{2020ApJ...892..147S}}
\end{threeparttable}
\end{table*} 

\subsection*{Distances and Galactocentric radii}

We derived the heliocentric distances ($d$) and Galactocentric radii ($R_{\rm gc}$) of the Galactic molecular clouds using the code provided by the BeSSeL survey \cite{2019ApJ...885..131R}. The Galactic rotation curve and spiral arm models were produced by measuring the trigonometric parallaxes and proper motions of $\sim 200$ molecular masers from the BeSSeL and the Japanese VLBI (very-long-baseline interferometry) Exploration of Radio Astrometry project. All Galactic molecular clouds studied in this work are close to the Galactic plane (Galactic latitude $|b|<4^{\circ}$), so the heights from the Galactic plane were negligible. The Galactocentric radius is, therefore, given by
\begin{equation}
R_{\rm gc}=\sqrt{R_{\rm gc,\; \odot}^{2}+d^{2}-2R_{\rm gc,\; \odot}d{\rm cos}(l)},
\end{equation}
where $R_{\rm gc, \odot}=8.15$ kpc is the Galactocentric radius of the Sun \cite{2019ApJ...885..131R}.

The inputs for the code are the Galactic longitude ($l$), Galactic latitude ($b$) and the local standard of rest velocity ($V_{\rm LSR}$). The code calculates the probability density function (PDF) of cloud distance through a Bayesian approach (`Data Availability'), from which the heliocentric distances ($d$) and their uncertainties were inferred. For each cloud, we used the most probable distance and the associated error given by the Bayesian inference. 

The PDF for the cloud distance was calculated by multiplying PDFs of (1) the association with a spiral arm model, (2) the kinematic distance from the rotation curve and (3) the vicinity to parallax sources. Each of these was weighted to construct the final PDF. However, because the spiral arm model in the far outer Galaxy ($R_{\rm gc}\gtrsim15$ kpc) was poorly constrained \cite{2019ApJ...885..131R}, we excluded the spiral arm model for all Galactic molecular clouds. Our results were not influenced, regardless of whether the spiral arm models were implemented. For most outer-disk clouds, there were no nearby parallax sources. Consequently, we adopted their kinematic distances based on the Galactic rotation curve.

For molecular clouds in the LMC, the SMC and NGC~6822, we adopted $d_{\rm LMC}=49.59\pm0.63$ kpc (ref.~\citen{2019Natur.567..200P}), $d_{\rm SMC}=62.44\pm1.28$ kpc (ref.~\citen{2020ApJ...904...13G}), and $d_{\rm NGC~6822}=474\pm13$ kpc (ref.~\citen{2014ApJ...794..107R}), respectively.

\subsection*{H$_{2}$ surface density}\label{SurfaceDensity}

We first generated source masks for each spectral cube using the source-finding algorithm embedded in \textsc{$^{\rm 3D}$Barolo} (ref.~\citen{2012MNRAS.421.3242W, 2015MNRAS.451.3021D}). For molecular clouds in the Galactic outer disk as well as the LMC, the SMC and NGC~6822, we set \texttt{MASK=Search}, \texttt{SNRCUT=3} and \texttt{GROWTHCUT=2.5}. Using this setting, signals in the line cube with $S/N>3$ were iteratively identified as real emission. Then, the source mask was grown to include surrounding signals with $S/N>2.5$. The data from the GRS for molecular clouds do not have enough line-free channels to estimate the spectral noise due to severe cloud blending. Therefore, we set \texttt{THRESHOLD=0.39} and \texttt{GROWTHTHRESHOLD=0.33}. With this setting, the algorithm searches for values larger than 0.39 K ($S/N=3$ in antenna temperature ($T_{\rm A}^{*}$)) and growth to 0.33 K ($S/N=2.5$ in $T_{\rm A}^{*}$) \cite{2006ApJS..163..145J} in the line cube.

We calculated the surface density of $^{13}$CO ($N_{\rm ^{13}CO}$) under the assumption of local thermodynamic equilibrium (LTE). For molecular clouds with complementary $^{12}$CO data, we estimated the excitation temperature ($T_{\rm ex}$) by assuming (1) that the low-$J$ $^{13}$CO lines share the same $T_{\rm ex}$ with the $^{12}$CO lines and (2) that the $^{12}$CO lines are optically thick. Therefore, $T_{\rm ex}$ is related to the peak main beam temperature of the $^{12}$CO line ($T_{\rm 12,pk}$) through
\begin{equation}
J(T_{\rm ex},\nu) = f_{\rm bm}^{-1}T_{\rm 12,pk}+J(T_{\rm bg},\nu),
\end{equation}
where $\nu$ is the rest frequency of the line. The background temperature ($T_{\rm bg}$) was taken as the brightness temperature of the cosmic microwave background ($T_{\rm CMB}=2.73$ K). The beam filling factor ($f_{\rm bm}$) was assumed to be unity. $J(T,\nu)$ is the Rayleigh-Jeans Equivalent Temperature \cite{2015PASP..127..266M}:
\begin{equation}
    J(T,\nu)\equiv\frac{h\nu/k}{{\rm exp}(h\nu/kT)-1}.
\end{equation}
where $k$ is the Boltzmann constant and $h$ is the Planck constant. When no complementary $^{12}$CO data were available, we took $T_{\rm ex}=15$ K, which is the typical kinetic temperature ($T_{\rm k}$) of infrared dark clouds \cite{2017A&A...598A..30T,2013A&A...552A..40C} and Planck Galactic cold clumps\cite{2022ApJS..258...17F}.

The optical depth of $^{13}$CO ($\tau_{13}$) of each voxel in the line cube was solved by the radiative transfer equation:
\begin{equation}
\label{equation:tau13}
\tau_{13}=-{\rm ln}\left\{1-\frac{T_{13}}{f_{\rm bm}[J(T_{\rm ex},\nu)-J(T_{\rm bg},\nu)]}\right\},
\end{equation} 
where $T_{\rm 13}$ is the main beam temperature of the $^{13}$CO line. Therefore, $N_{\rm ^{13}CO}$ was found by integrating $\tau_{13}$ along the spectral axis, that is the velocity ($v$) \cite{1991ApJ...374..540G}: 
\begin{equation}
\label{eq:N13}
N_{\rm ^{13}CO} = \frac{3k}{8\pi^{3}B\mu^{2}}{\rm exp}\left[\frac{hBJ_l(J_l+1)}{kT_{\rm ex}}\right] \times \frac{T_{\rm ex}+hB/3k}{1-{\rm exp}(-h\nu/kT_{\rm ex})}\times \frac{\int \tau_{13}dv}{(J_l+1)},
\end{equation}
where $B$ ($\sim55.1$ GHz) is the rotational constant, $\mu$ ($\sim0.112$ D) is the dipole moment and $J_{l}$ is the lower energy level of the $^{13}$CO transition (for $^{13}$CO $J=1\rightarrow0$, $J_{l}$=0; for $^{13}$CO $J=2\rightarrow1$, $J_{l}$=1).

We converted $N_{\rm ^{13}CO}$ to $N_{\rm H_{2}}$ through the abundance ratios of $N_{\rm ^{12}CO}/N_{\rm ^{13}CO}$ (hereafter $^{12}$CO/$^{13}$CO) and $N_{\rm H_{2}}/N_{\rm ^{12}CO}$ (hereafter H$_{2}$/$^{12}$CO) for the Galactic molecular clouds:
\begin{equation}
    N_{\rm H_{2}} = N_{\rm ^{13}CO}\times {\rm ^{12}CO/^{13}CO\times H_{2}/^{12}CO}.
    \label{eq:NH2}
\end{equation} 
The underlying assumptions are as follows: (1) $\rm ^{12}CO/^{13}CO$ is represented by the $^{12}$C-to-$^{13}$C isotopic ratio ($^{12}$C/$^{13}$C$=5.87R_{\rm gc}+13.25$)\cite{2020A&A...640A.125J} and (2) H$_{2}$/$^{12}$CO is inversely proportional to the gas-phase oxygen abundance (O/H$\propto10^{-0.044(R_{\rm gc}-R_{\rm gc,\odot})}$)\cite{2022MNRAS.510.4436M}. The H$_{2}$/$^{12}$CO value at $R_{\rm gc}=R_{\rm gc, \odot}$ is $6.0\times10^{3}$, measured by absorption lines in nearby clouds against background stars\cite{2017ApJ...838...66L}.

We assumed $\rm H_{2}/^{13}CO=3\times10^{6}$ (ref.~\citen{2014ApJ...796..123F, 2019ApJ...885...50W}) for molecular clouds in the LMC. We adopted $\rm H_{2}/^{13}CO=7.5\times10^{6}$ for the SMC and NGC~6822, which was scaled from the LMC value through their metallicity ratio.

\subsection*{Physical properties of molecular clouds}

In this work, the equivalent radius ($R_{\rm cloud}$), the velocity dispersion ($\sigma_{\rm v}$) and the molecular gas mass ($M_{\rm mol}$) of molecular clouds were all measured within the half-peak isophote of $N_{\rm H_{2}}$. This enabled a consistent comparison among observations with different angular resolutions and sensitivities (`Possible bias in measuring cloud properties'). The measured physical quantities for the Galactic molecular clouds are presented in Supplementary Tables~\ref{tab: Supp_Table1} and \ref{tab: Supp_Table2}.

\noindent $\bullet$ $R_{\rm cloud}$: The equivalent angular radius ($r_{\rm cloud}$) of a cloud was calculated by deconvolving the telescope beam from the observed angular area ($A$; within the half-peak isophote of $N_{\rm H_{2}}$):
\begin{equation}
    r_{\rm cloud} = \sqrt{\frac{A}{\pi}-\frac{\theta_{\rm beam}^{2}}{4}}.
    \label{eq:rcloud}
\end{equation}

Theoretically, the fractional uncertainty of $r_{\rm cloud}$ is inversely proportional to the $S/N$ of the peak intensity\cite{1997PASP..109..166C}. Most clouds studied in this work have a peak $S/N$ much larger than 10. Therefore, we adopted a conservative uncertainty of $10\%$ for $r_{\rm cloud}$ (`Possible bias in measuring cloud properties') to include any unforeseen errors. The cloud physical radius is then given by $R_{\rm cloud} = r_{\rm cloud}d$.

\noindent $\bullet$ $\sigma_{\rm v}$: The intensity-weighted velocity dispersion ($\sigma_{\rm v}$) is given by 
\begin{equation}
        \sigma_{\rm v} = \sqrt{\sigma_{\rm v,obs}^{2}-\sigma_{\rm v,ins}^{2}},
        \label{equation:sigma_v}
\end{equation}
where $\sigma_{\rm v,obs} = \sqrt{\sum T_{\rm i}(v_{\rm i}-\bar v)^2/\sum T_{\rm i}}$ is the observed velocity dispersion and $\sigma_{\rm v,ins}=\Delta V_{\rm ins}/{\rm 2\sqrt{2ln(2)}}$ is the velocity dispersion led by the instrumental spectral broadening. $\bar v = \sum T_{\rm i}v_{\rm i}/{\sum T_{\rm i}}$ is the intensity-weighted mean velocity. $\Delta V_{\rm ins}$ is approximately the channel width. We also applied an uncertainty of $10\%$ for $\sigma_{\rm v}$.

\noindent $\bullet$ $M_{\rm mol}$: The $M_{\rm mol}$ of a molecular cloud at a distance of $d$ is
\begin{equation}
    M_{\rm mol}=\mu m_{\rm H_{2}} d^{2}\int_{A} N_{\rm H_{2}}\delta x \delta y,
\end{equation}
where $m_{\rm H_{2}}\approx3.347115\times10^{-24}$ g is the mass of an \Htwo\ molecule, $\delta x$ and $\delta y$ are the pixel angular sizes and $\mu\approx1.36$ is the mean molecular weight considering the mass of helium \cite{knapen2017outskirts}. We adopted an uncertainty of $20\%$ for $M_{\rm mol}$, which is a conservative estimate for the flux calibration error for millimetre-wave observations. Systematic errors are not involved in error propagation.

The mean mass surface density ($\Sigma_{\rm mol}$) is
\begin{equation}
    \Sigma_{\rm mol} = \frac{M_{\rm mol}}{\pi R_{\rm cloud}^{2}}.
\end{equation}

For molecular clouds in DDO~70, $\Sigma_{\rm mol}$ is related to the cloud-boundary surface density ($\Sigma_{\rm limit}=756\pm468\,M_{\odot}\,\rm pc^{-2}$, projection on the two-dimensional sky)\cite{2020ApJ...892..147S} by $\Sigma_{\rm mol}=\frac{2}{-k+3}\Sigma_{\rm limit}$, for a radial density profile $\rho\propto r^{-k}$. We adopt $k=1.8$ for typical star forming clumps in the Milky Way \cite{2002ApJS..143..469M}.

\subsection*{The virial parameter}

The virial parameter ($\alpha_{\rm vir}$) is defined as\cite{1992ApJ...395..140B}:
\begin{equation}
\alpha_{\rm vir} = \frac{5\sigma_{\rm v}^{2}R_{\rm cloud}}{GM_{\rm mol}}=a\frac{2E_{\rm k}}{|E_{\rm g}|},
\label{eq: alphavir}
\end{equation}
where $G$ is the gravitational constant. For a uniform-spherical cloud ($a=1$)\cite{1992ApJ...395..140B}, if the contributions of both the external pressure and the magnetic field are negligible, virial equilibrium ($2E_{\rm k}+E_{\rm g}=0$) and energy equipartition ($E_{\rm k}+E_{\rm g}=0$) lead to $\alpha_{\rm vir}=1$ and $\alpha_{\rm vir}=2$, respectively. Even though virial equilibrium and energy equipartition are conceptually different\cite{2006MNRAS.372..443B}, a factor of two in $\alpha_{\rm vir}$ is hard to measure due to the systematic uncertainties. Consequently, we could not distinguish them in this work. Molecular clouds with $\alpha_{\rm vir}>1$ and $\alpha_{\rm vir}<1$ are defined as supervirial and subvirial, respectively. 

\subsection*{Systematic errors in $M_{\rm mol}$ and \alphavir}

The uncertainties in $M_{\rm mol}$ and \alphavir\ come from the uncertainties in the cloud distance ($d$), the excitation conditions of $^{13}$CO and the abundance ratios. Among the parameters used to calculate \alphavir\ (Eq.~\ref{eq: alphavir}), both $R_{\rm cloud}$ and $M_{\rm mol}$ depend on the cloud distance. $R_{\rm cloud}$ scales with $d$. $M_{\rm mol}$ scales with both $d^2$ and the \Htwo/\thco\ abundance ratio. The \Htwo/\thco\ abundance ratio was estimated through the Galactic $^{12}$C/$^{13}$C gradient and the O/H gradient, and is, therefore, a function of $R_{\rm gc}$. Therefore, \alphavir $\propto d^{-1}({\rm H_{2}/^{13}CO})^{-1}$.

The distances of the Galactic outer-disk molecular clouds are well constrained with a fractional uncertainty of $\lesssim20\%$. For clouds at $R_{\rm gc}>15$ kpc, the distance uncertainty contributes $\lesssim40\%$ of the uncertainty in $M_{\rm mol}$ and $\lesssim20\%$ of the uncertainty in \alphavir.

In the following, we assess the systematic errors arising from uncertainties in $^{13}$CO excitation conditions and abundance ratios.

\begin{itemize}

\item {The excitation of $^{13}$CO.} Depending on the specific excitation conditions\cite{2009ApJ...699.1092H}, assuming local thermodynamic equilibrium can deviate $N_{\rm ^{13}CO}$ from its intrinsic value. For the typical mean volume density of \Htwo\ ($n_{\rm H_2}$) of the molecular clouds studied in this work ($\sim10^{3}-10^{4}$ cm$^{-3}$), the deviation from the theoretical mass was $\lesssim40\%$. The assumption of $T_{\rm ex}=15$ K may overestimate $N_{\rm ^{13}CO}$ by $\lesssim20\%$ for the $J=1\rightarrow0$ transition (Supplementary Fig.~\ref{fig: Supp_Fig1}a). The overestimation is negligible for the $J=2\rightarrow1$ transition (Supplementary Fig.~\ref{fig: Supp_Fig1}b). Therefore, the overall overestimation due to the assumptions on $^{13}$CO excitation is within 60\%.

\begin{figure}[t!]
\includegraphics[width=0.5\linewidth]{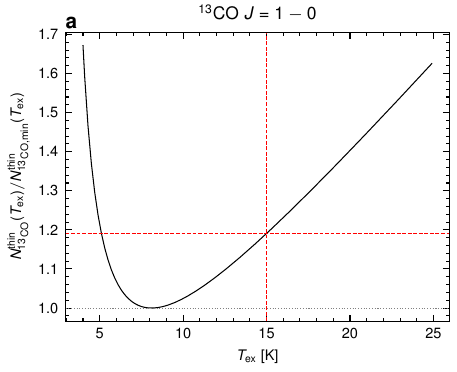}
\includegraphics[width=0.5\linewidth]{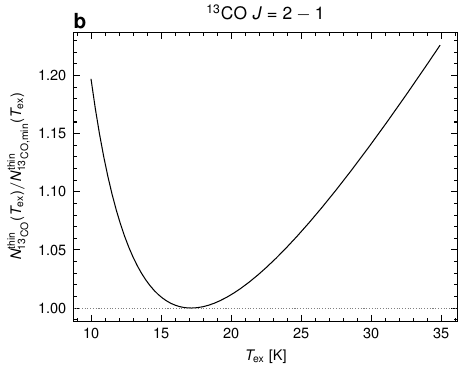}
\caption{\textbf{Effect of $T_{\rm ex}$ on $N_{\rm ^{13}CO}$ estimate}. We assume that the $^{13}$CO lines are in local thermodynamic equilibrium and optically thin. The curves are scaled with respect to their minimum values. \textbf{a,} For $^{13}$CO $J=1-0$. \textbf{b,} For $^{13}$CO $J=2-1$.}
\label{fig: Supp_Fig1}
\end{figure}

\item {The abundance ratios.} Not only are the $^{12}$C/$^{13}$C ratios still poorly measured at $R_{\rm gc}>12$ kpc (ref.~\citen{1996A&AS..119..439W,2005ApJ...634.1126M,2020A&A...640A.125J,2024MNRAS.527.8151S}) but the Galactic gas-phase O/H versus $R_{\rm gc}$ gradient is also still less constrained at $R_{\rm gc}>18$ kpc (ref.~\citen{2022MNRAS.510.4436M}). Depending on the depth into molecular clouds, the molecular ($^{12}$CO/$^{13}$CO) abundance ratio can be lower than the isotopic ($^{12}$C/$^{13}$C) abundance ratio due to isotopic-selective chemical reactions\cite{2014MNRAS.445.4055S}. This may have led to an overestimation of $N_{\rm ^{12}CO}$ as inferred from $N_{\rm ^{13}CO}$ by up to $50-60\%$ (ref.~\citen{2014MNRAS.445.4055S}). The H$_{2}$/$^{12}$CO ratios measured in the solar neighbourhood ($R_{\rm gc,\odot}$) also vary by a factor of two among different studies \cite{1978ApJS...37..407D,1982ApJ...262..590F,2017ApJ...838...66L}. Despite these uncertainties, the dependencies of $^{12}$CO/$^{13}$CO and H$_{2}$/$^{12}$CO on $R_{\rm gc}$ (or $Z$) are natural expectations of Galactic chemical evolution models (Supplementary Fig.~\ref{fig: Supp_Fig2})\cite{2019MNRAS.490.2838R} and astrochemistry\cite{2012MNRAS.426..377G}.  Supplementary Fig.~\ref{fig: Supp_Fig3} shows a test under an extreme condition, where $M_{\rm mol}$ is calculated with a constant \Htwo/\thco\ abundance ratio (to simulate what would happen if we were to neglect any abundance gradients). In this case, the \alphavir\ trends become flatter. Given both the observational and theoretical evidence for the existence of gradients in both $^{12}$C/$^{13}$C and O/H with $R_{\rm gc}$, the \alphavir\ trends can be considered as solid, albeit the uncertainties in the gradient slopes.

\begin{figure}[t!]
\includegraphics[width=\linewidth]{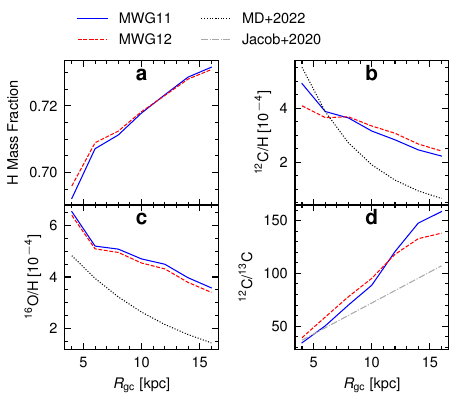}
\caption{\textbf{Galactic radial gradients of elemental abundances.} In each panel, the blue solid and the red dashed lines show the $R_{\rm gc}$-gradients predicted by the Galactic Chemical Evolution models \cite{2019MNRAS.490.2838R}. The black dotted lines show the C/H-$R_{\rm gc}$ gradient in \textbf{b} and the O/H-$R_{\rm gc}$ gradient in \textbf{c} from observations\cite{2022MNRAS.510.4436M}. The grey dash-dotted line in \textbf{d} shows the observed $^{12}$C/$^{13}$C-$R_{\rm gc}$ gradient\cite{2020A&A...640A.125J}.}
\label{fig: Supp_Fig2}
\end{figure}

\begin{figure}[h!]
\includegraphics[width=1\linewidth]{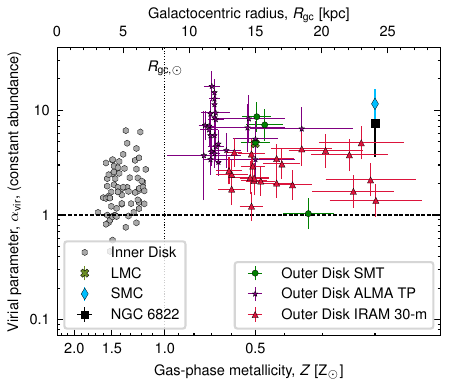}
\caption{\textbf{Examining the \alphavir\ trends under an extreme assumption.} Instead of considering an H$_{\rm 2}$/$^{13}$CO ratio that varies with $R_{\rm gc}$ (or $Z$), we convert $N_{\rm ^{13}CO}$ to $N_{\rm H_{2}}$ using a constant H$_{\rm 2}$/$^{13}$CO value (set by its value at $R_{\rm gc,\odot}$). Molecular clouds in DDO~70 are not shown in this plot because their cloud masses are measured following a different approach \cite{2020ApJ...892..147S}. Data are presented as measured values with 1-$\sigma$ uncertainties.}
\label{fig: Supp_Fig3}
\end{figure}

\end{itemize}

\subsection*{Possible bias in measuring cloud properties}\label{MeasurementBias}

Limited angular resolution and observational sensitivity may also have biased the measurement of $R_{\rm cloud}$ and $M_{\rm cloud}$. To examine to what extent these may have influenced our results, we generated a set of mock clouds to mimic real observations. First, we generated two-dimensional Gaussian models as the intrinsic cloud emissions. We then convolved the Gaussian models with the telescope beam and added Gaussian noise to the convolved models.

The peak values of the Gaussian models were scaled such that the final convolved maps have a peak $S/N=10$, which is approximately the worst $S/N$ of our sample of Galactic clouds. The major axes of the Gaussian models were set from 4$''$ to 100 $''$ (with a spacing of 0.5$''$) and the minor-to-major axis ratios (aspect ratios) were set from 0.2 to 1.0 (with a spacing of 0.2). We used a telescope beam of 23.5$''$ and a pixel size of 4$''$, the same as for our IRAM 30 m telescope observations. Similarly, we measured the cloud equivalent radii and fluxes within the half-peak isophote of the convolved models.

\begin{figure}[h!]
\includegraphics[width=0.5\linewidth]{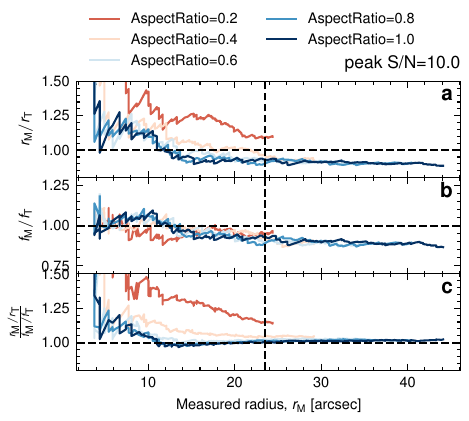}
\includegraphics[width=0.5\linewidth]{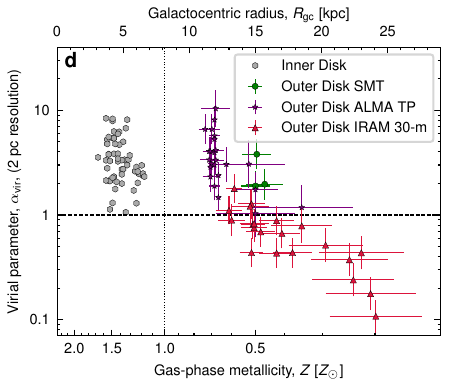}
\caption{\textbf{Examining measurement bias led by the spatial resolution.} The effect of angular resolution and sensitivity on measuring the cloud angular radius $r$ (panel \textbf{a}), the emission flux $f$ (panel \textbf{b}), and the virial parameter $\propto r/f$ (panel \textbf{c}). The subscripts `M' and `T' refer to the Measured value and the True value, respectively. \textbf{d,} Before measuring the cloud physical quantities, all line cubes of the Galactic molecular clouds are spatially smoothed to a uniform physical resolution of 2 pc. Data are presented as measured values with 1-$\sigma$ uncertainties.}
\label{fig: Supp_Fig4}
\end{figure}

Supplementary Fig.~\ref{fig: Supp_Fig4}a shows the ratio between the measured radius ($r_{\rm M}$) and its true value ($r_{\rm T}$) as a function of $r_{\rm M}$. For clouds with an aspect ratio $\ge 0.4$ (roundish clouds) and $r_{\rm M}\gtrsim10''$ (clouds large enough to be marginally resolved), the difference between $r_{\rm M}$ and $r_{\rm T}$ is within 10\%, consistent with the theoretical prediction\cite{1997PASP..109..166C}. In other cases, that is, elongated clouds and small clouds, the cloud radii are overestimated because the minor axes in the convolved models are not well sampled by the image pixel.

Supplementary Fig.~\ref{fig: Supp_Fig4}b shows the ratio between the measured flux ($f_{\rm M}$) and its true value ($f_{\rm T}$) as a function of $r_{\rm M}$. In all cases, the difference between $f_{\rm M}$ and $f_{\rm T}$ is within 10\%. Given that the flux ratio is representative of the mass ratio (Eq.~\ref{eq:N13}) for optically thin, low-$J$ $^{13}$CO lines, the cloud masses in this work are insensitive to measurement bias.

Consequently, $\frac{r_{\rm M}/r_{\rm T}}{f_{\rm M}/f_{\rm T}}$ measures the ratio between the measured \alphavir\ and its true value. Supplementary Fig.~\ref{fig: Supp_Fig4}c shows $\frac{r_{\rm M}/r_{\rm T}}{f_{\rm M}/f_{\rm T}}$ as a function of $r_{\rm M}$. The measured \alphavir\ was almost identical to its true value unless the cloud was too small to be sampled by the image pixel, where the \alphavir\ is overestimated. Given that most of the clouds in this work have peak $S/N>10$ and $r_{\rm M}>10''$, the angular resolution and the sensitivity do not bias the measurement of \alphavir.

Given the wide range of cloud distances in the Milky Way, a specific telescope beam corresponds to different physical resolutions. We, thus, further tested to see whether this biased our measurements of the cloud \alphavir. First, we smoothed all line cubes for the Galactic molecular clouds to the same physical resolution of 2 pc, where the smoothing kernels depend on the cloud distances. A few clouds were excluded due to their small (physical) mapping areas. Then, we followed the same procedures to measure the cloud \alphavir. Supplementary Fig.~\ref{fig: Supp_Fig4}d shows that the $\alpha_{\rm vir}$ trends are not affected by the downgraded physical resolution.

\subsection*{Selection bias}

A larger heliocentric distance could bias the sample to brighter (and, thus, more massive) molecular clouds, which tend to have smaller $\alpha_{\rm vir}$ in observations\cite{2013ApJ...779..185K}. 

\begin{figure}[h!]
\includegraphics[width=1\linewidth]{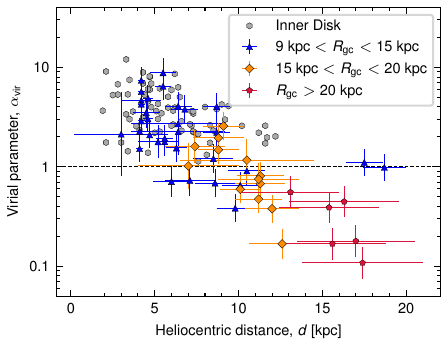}
\caption{\textbf{Examining bias led by the heliocentric distance.} Variation of $\alpha_{\rm vir}$ with $d$. The outer disk clouds are divided into three $R_{\rm gc}$ bins. Data are presented as measured values with 1-$\sigma$ uncertainties.} 
\label{fig: Supp_Fig5}
\end{figure}

We examined whether the decreasing $\alpha_{\rm vir}$ trends were biased by such selection effects by comparing molecular clouds with different $R_{\rm gc}$ but similar heliocentric distances (Supplementary Fig.~\ref{fig: Supp_Fig5}). At the distance range $\sim 15-20$ kpc, the sample contains two molecular clouds at $R_{\rm gc}\approx13$ kpc (G37.350 and G44.8 in the first Galactic quadrant) and five molecular clouds at $R_{\rm gc}>20$ kpc (SUN15\_56, SUN15\_57, SUN15\_53, SUN15\_55, and SUN15\_59 in the second Galactic quadrant). The clouds at $R_{\rm gc}>$ 20 kpc have at least $\times$5 smaller \alphavir\ than those at $R_{\rm gc}\approx13$ kpc.

\begin{figure}[h!]
\includegraphics[width=0.5\linewidth]{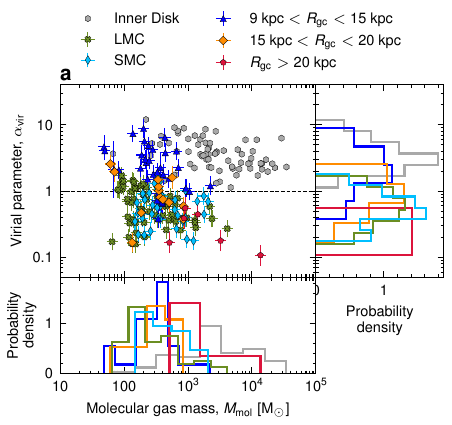}
\includegraphics[width=0.5\linewidth]{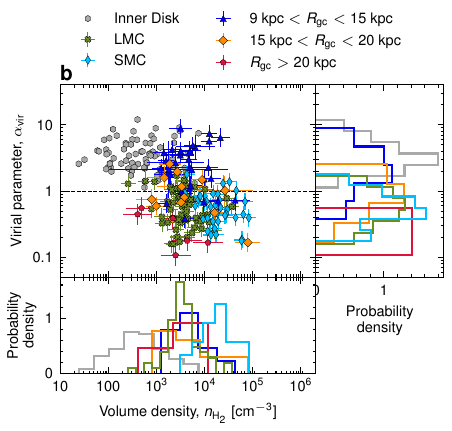}
\caption{\textbf{Examining bias led by the molecular gas mass and the mean volume density.} \textbf{a,} The upper left panel shows the variation of $\alpha_{\rm vir}$ with $M_{\rm mol}$. All molecular clouds in the LMC and the SMC are plotted. The upper right and the lower left panels show the probability density distribution of \alphavir\ and $M_{\rm mol}$ for different cloud samples. \textbf{b,} Variation of $\alpha_{\rm vir}$ with $n_{\rm H_{2}}$. Data are presented as measured values with 1-$\sigma$ uncertainties.}
\label{fig: Supp_Fig6}
\end{figure}

Supplementary Fig.~\ref{fig: Supp_Fig6}a,b show that for each $R_{\rm gc}$ bin, $\alpha_{\rm vir}$ barely varies with $M_{\rm mol}$ or $n_{\rm H_{2}}$. At comparable $M_{\rm mol}$, $\alpha_{\rm vir}$ becomes smaller for larger $R_{\rm gc}$ (reminiscent of predictions by analytic models on Giant Molecular Cloud scales \cite{2018ApJ...854..100M}). Therefore, the distance bias to the \alphavir\ trends is not important.

\subsection*{The general virial theorem}

From the momentum equation, the general Virial Theorem in the presence of external pressure and magnetic fields can be derived \cite{spitzer2008physical,1989ApJ...338..178E,1992ApJ...395..140B} as:
\begin{equation}
\frac{\ddot{I}}{2}=2(E_{\rm k}-E_{\rm k,0})+E_{\rm m}+E_{\rm g},
\label{eq:virial}
\end{equation}
where $E_{\rm k}=\frac{3}{2}M_{\rm mol}\sigma_{\rm v}^{2}$ is the kinetic energy and $E_{\rm g}=-a\frac{3GM_{\rm mol}^{2}}{5R_{\rm cloud}}$ is the self-gravitational energy ($a$ is of order unity and accounts for the non-sphericity and the non-uniformity of a molecular cloud). $E_{\rm k,0}=\frac{3P_{\rm e}V}{2}$ and $E_{\rm m}=\frac{1}{8\pi}\int(B^{2}-B^{2}_{\rm e})dV$ account for the external pressure ($P_{\rm e}$) and the magnetic field ($B$), surrounding and within the cloud volume ($V$), respectively. $B_{\rm e}$ is the magnetic field outside the cloud. 

For the general virial equilibrium, the second-derivative of the moment of inertia ($\ddot{I}$) equals zero unless tidal fields are dominant within a cloud volume, which may be the case for clouds in the Galactic Centre. In Eq.~\ref{eq:virial}, only the volumetric $\frac{1}{8\pi} \int B^2 dV$ term inside $E_{\rm m}$ has the sign that allows it to serve as another support in the dynamic state of the cloud besides $E_{\rm k}$.

\begin{figure}[h!]
 \includegraphics[width=\linewidth]{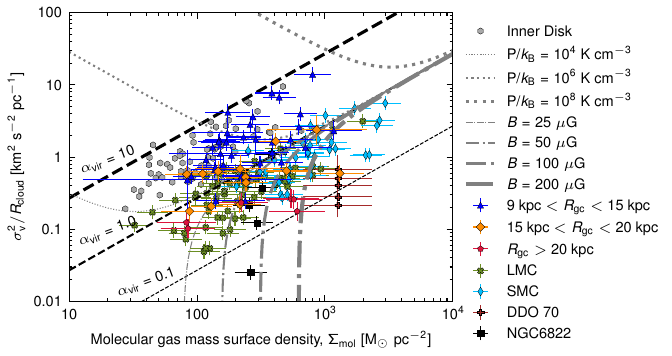}
\caption{\textbf{Dynamic analysis based on the general virial theorem.} The black dashed lines show $\alpha_{\rm vir}=0.1$, 1, and 10. Molecular clouds with $\alpha_{\rm vir}>1$ and $\alpha_{\rm vir}<1$ are super-virial and sub-virial, respectively. Assuming the general virial theorem (see Methods), the super-virial molecular clouds should be bound by external pressure ($P_{\rm e}$) while the sub-virial molecular clouds should be supported by magnetic fields ($B$). The grey dotted lines correspond to $P_{\rm e}=$ 10$^{4}$, 10$^{6}$, and 10$^{8}$ $k_{\rm B}$ K cm$^{-3}$. The grey dash-dotted lines correspond to $B=$ 25, 50, 100, and 200 $\mu$G, from thin to thick. Data are presented as measured values with 1-$\sigma$ uncertainties.} 
\label{fig: ED_Fig1}
\end{figure}

For the supervirial molecular clouds, the support from the $B$ field is often negligible due to the strong turbulence. If these clouds are in general virial-equilibrium, they are most probably bound by $P_{\rm e}$ (lines of constant pressures in Supplementary Fig.~\ref{fig: ED_Fig1})\cite{1989ApJ...338..178E,1992ApJ...395..140B,2011MNRAS.416..710F}:
\begin{equation}
    \frac{\sigma_{\rm v}^{2}}{R_{\rm cloud}}=\frac{\pi G \Sigma_{\rm mol}}{5}+\frac{4P_{\rm e}}{3\Sigma_{\rm mol}}.
    \label{eq:pressure-virial}
\end{equation}

Supervirial $\alpha_{\rm vir}$ values are often deduced for molecular clouds in the centres of galaxies and are thought to be responsible for the lower $X_{\rm CO}$ values found for them \cite{2014A&A...568A.122Z,1999ApJ...516..114P}.

For the subvirial clouds, the $B$ field starts to dominate the support against cloud self-gravity. The supporting force from the $B$ field is indirectly exerted on the neutral molecular gas through ion-neutral collisions. Given the representative volume density ($n_{\rm H_{2}}\approx10^{3}-10^{4}$ cm$^{-3}$) of the outer-disk clouds and a typical cosmic-ray ionization rate ($\zeta_{\rm H_{2}}=10^{-17}$ s$^{-1}$) \cite{1979ApJ...232..729E,2023pcsf.conf..237P}, the typical ionization fraction ($x_{\rm e}=n_{e}/n_{\rm H}$)\cite{1998ApJ...503..689W} is $\sim10^{-7}$. This results in a magnetohydrodynamic cutoff wavelength of $\lesssim0.01$ pc (ref.~\cite{1998ApJ...503..689W}). Therefore, for the typical scale ($\sim1$ pc) of the outer-disk molecular clouds, the magnetic fields are well coupled with neutral particles. Neglecting the minor contribution from $P_{\rm e}$ in the outer Galaxy, we can derive (lines with constant $B$ fields in Supplementary Fig.~\ref{fig: ED_Fig1}):
\begin{equation}
    \frac{\sigma_{\rm v}^{2}}{R_{\rm cloud}}=\frac{\pi G \Sigma_{\rm mol}}{5}-\frac{B^{2}}{18\pi\Sigma_{\rm mol}}.
    \label{eq:B-field-virial}
\end{equation} 
Here we neglect the external magnetic field $B_{\rm e}$ term as it is small ($\sim$10 $\mu$G)\cite{2012ARA&A..50...29C} compared to the $B$ field required to support the subvirial clouds ($\sim$ 100 $\mu$G), as we show below.

\begin{figure}[h!]
\includegraphics[width=1.0\linewidth]{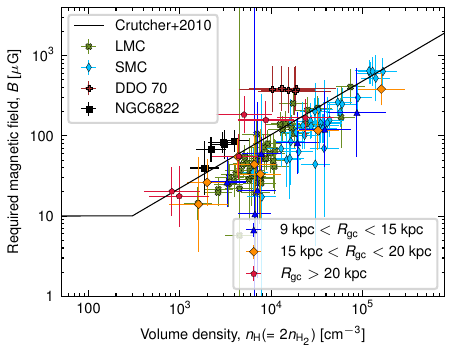}
\caption{\textbf{Variation of the magnetic field strength required to support the sub-virial clouds as a function of the gas volume density.} In molecular clouds, the gas volume density is $n_{\rm H}=2n_{\rm H_{2}}$. The $B$-field required for supporting the sub-virial molecular clouds is calculated from Eq.~\ref{eq:B-field-virial}, which is derived from the general virial theorem. The $B-n_{\rm H}$ relation benchmarked through observations in the Solar neighborhood and the inner Galaxy \cite{2010ApJ...725..466C} is overlaid as the black solid curve. Data are presented as measured values with 1-$\sigma$ uncertainties.}
\label{fig: ED_Fig2}
\end{figure}

Supplementary Fig.~\ref{fig: ED_Fig2} compares the $B$ field required to support the subvirial clouds with the expectations from the $B$ versus $n_{\rm H}$ relation ($n_{\rm H}=2n_{\rm H_{2}}$) \cite{2010ApJ...725..466C,2022Natur.601...49C}. This relation was established in the solar neighbourhood and the inner Galaxy by measuring Zeeman effects. A classical explanation of this relation is that the $B$-field line is coupled with the molecular gas through ion-neutral coupling \cite{lequeux2004interstellar} (which is solid at $>0.01$ pc scales). In denser regions ($n_{\rm H}>300$ cm$^{-3}$), the field lines are squeezed by the self-gravity of a cloud, so the $B$ field is enhanced \cite{2010ApJ...725..466C}. There are also other explanations \cite{2023ApJ...946L..46C}.

Supplementary Fig.~\ref{fig: ED_Fig2} suggests that a similar $B$-field strength to those in the inner Galaxy is sufficient for supporting the subvirial clouds in the outer Galaxy. In nearby Milky Way-like spiral galaxies, the energy density of the $B$ field (proportional to $B^2$) on kiloparsec scales varies only a little from the inner to the outer galactic disk (except for the innermost regions) \cite{2013MNRAS.433.1675B,2015A&A...578A..93B}. If this is also true for the Milky Way, we expect that the $B$ field in molecular clouds will be similar across the Galactic plane, provided that the mechanisms for enhancing the $B$ field with density are similar.

Possible deviations from a uniform-spherical cloud ($a>1$) would even further consolidate our results. This is because \alphavir\ defined by Eq.~\ref{eq: alphavir} overestimates $2E_{\rm k}/|E_{\rm g}|$ by a factor of $a$. Given the typical density profiles of Galactic star-forming clouds\cite{2002ApJS..143..469M}, $a\sim1.4\pm0.3$. To properly measure $a$, high-angular-resolution observations are needed\cite{2013ApJ...768L...5L}.

\subsection*{Contribution from cloud rotation and thermal motion in $E_{\rm k}$}

In addition to the random turbulent motion, cloud rotation and thermal motion can also contribute to the $\sigma_{\rm v}$ of a molecular cloud. To evaluate the effect of cloud rotation, we first fitted the rotational velocity along each LoS with a planar function:
\begin{equation}
v_{\rm rot}(x,y) = a(x-x_{0})+b(y-y_{0})+v_{0},
\end{equation}
where ($x_{0}$, $y_{0}$) is the centre coordinate and $v_{0}$ is the LoS velocity at ($x_{0}$, $y_{0}$). Then we subtracted the rotational velocity along each LoS in Eq.~\ref{equation:sigma_v} through
\begin{equation}
        \sigma_{\rm v,obs}^{\rm non-rot} = \sqrt{\frac{\sum T_{\rm i}(v_{\rm i}-\bar v - v_{\rm rot})^2}{\sum T_{\rm i}}}.
\end{equation}
The contribution of cloud rotation in $E_{\rm k}$ is $\lesssim$ 10$\%$.

\begin{figure}[t!]
\includegraphics[width=1\linewidth]{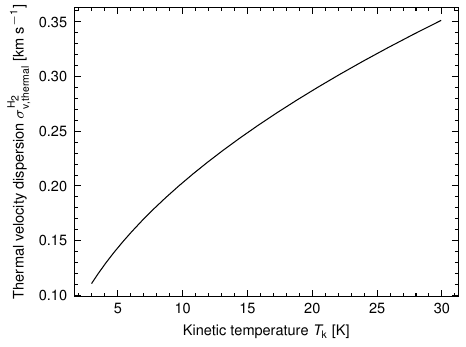}
\caption{\textbf{Thermal broadening of H$_{2}$.} Variation of thermal velocity dispersion of H$_{2}$ ($\sigma_{\rm v,thermal}^{\rm H_{2}}$) with kinetic temperature ($T_{\rm k}$).}
\label{fig: Supp_Fig7}
\end{figure}

The thermal broadening of $^{13}$CO ($\sigma_{\rm v,thermal}^{\rm ^{13}CO}$) at $T_{\rm k}=15$ K is $\sim$ 0.07 $\rm km\;s^{-1}$, which is negligible in $\sigma_{\rm v}$. However, molecular clouds are composed mostly of H$_{2}$ molecules, which are lighter than $^{13}$CO molecules. Therefore, at the same temperature, the thermal velocity dispersion of \Htwo\ should be larger than that of \thco. For $T_{\rm k}=15$ K, $\sigma_{\rm v,thermal}^{\rm H_{2}} =$ 0.25 km s$^{-1}$ (Supplementary Fig.~\ref{fig: Supp_Fig7}). For the outer-disk molecular clouds, the smallest $\sigma_{\rm v}$ is $\sim$ 0.30 km s$^{-1}$. Therefore, for the most extreme cases, including the thermal motion in the total $E_{\rm k}$ would increase $\alpha_{\rm vir}$ by $\lesssim70\%$. 

\subsection*{Possible mechanisms for $B$-field-dominated cloud dynamics at low metallicities}

Our results reveal that the dynamics of metal-poor clouds may be dominated by the $B$ field. However, the detailed underlying physical mechanisms are still unclear. Here we propose two speculations that may play a role. 

First, as was revealed by magnetohydrodynamic simulations \cite{2023ApJ...954...38K}, the velocity dispersion of the cold neutral medium decreases towards low metallicity. As molecular clouds should inherit the dynamic properties from their parental ambient atomic gas \cite{2022MNRAS.515.2822R}, they are also expected to have low turbulence levels in low-metallicity conditions. This is in line with Fig.~\ref{fig: Main_Fig2}b as well as the narrow CO linewidths found in the LMC and SMC \cite{2008ApJ...686..948B,2023ApJ...949...63O}. The cold neutral medium surrounding metal-poor molecular clouds, which has a higher fraction under low-metallicity conditions, may allow turbulence to dissipate significantly enough before entering the molecular clouds. In this case, the $B$ field would take over the role of supporting clouds against self-gravity, resulting in a low \alphavir\ observed in metal-poor molecular clouds. 

Second, the far-ultraviolet photons can penetrate deeper into molecular clouds in low-metallicity conditions, leading to a higher ionization fraction. This would allow stronger coupling between neutrals and ions regulated by the $B$ field\cite{lequeux2004interstellar} in such conditions. This could enhance the B-field contribution in supporting these clouds, thus decreasing \alphavir.

Apart from the above mechanisms, the \thco\ molecules could preferentially survive in regions with high column densities, under low-metallicity conditions. This may cause a bias towards the dense cores of molecular clouds, which are often found subvirial with high-volume-density gas tracers \cite{2013ApJ...779..185K,2024ApJ...974L...6W}. However, the low-metallicity clouds in this work have a typical size of $\sim1$ pc, which is larger than dense cores in the solar neighbourhood. This indicates that the $B$ field is already important on cloud scales under low metallicities.

\subsection*{Low turbulence injection and star-formation activities in the Galactic outer disk}

\begin{figure}[t!]
\includegraphics[width=\linewidth]{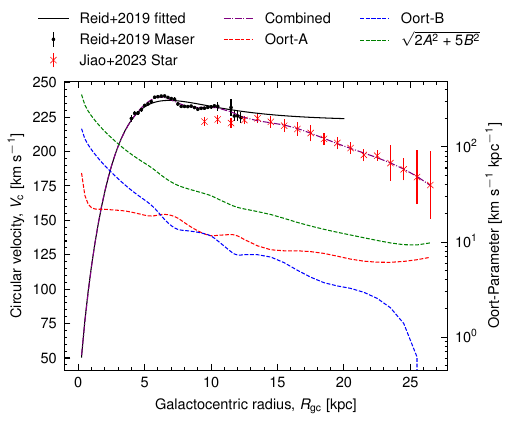}
\caption{\textbf{The Galactic rotation curve and the Oort parameters.} The Galactic radial profile of the Circular Velocity ($V_{\rm c}$) derived by different methods are shown as black solid line/dots\cite{2019ApJ...885..131R} and red errorbars\cite{2023A&A...678A.208J}). The combined rotation curve (purple dash-dotted line) is adopted to calculate the Oort-$A$ (red dashed line) and Oort-$B$ (blue dashed line) parameters \cite{2020MNRAS.496.5211A}, which measure the strength of shear and vorticity due to the Galactic rotation curve, respectively. The green dashed line shows the square root of the kinetic energy ($\propto2A^{2}+5B^{2}$) led by the Galactic differential rotation \cite{2020MNRAS.496.5211A}. Data are presented as measured values with 1-$\sigma$ uncertainties.}
\label{fig: Supp_Fig8}
\end{figure}

In the Milky Way, the turbulence-injection level is expected to decrease towards the outer Galaxy. Turbulence in molecular clouds can be driven by both global Galactic dynamics \cite{2006ApJ...638..191K,2018ApJ...854..100M,2022MNRAS.515.2822R} and local stellar feedback \cite{2006ApJ...641L.121T,2016ApJ...822...11P}. On galactic scales, both shear (Oort constant $A$) and vorticity (Oort constant $B$) induced by Galactic differential rotation decline with $R_{\rm gc}$ (Supplementary Fig.~\ref{fig: Supp_Fig8}), leading to less kinetic energy (proportional to $2A^{2}+5B^{2}$) injected by cloud-cloud collisions \cite{2020MNRAS.496.5211A}. On the other hand, the surface density of the Galactic star formation rate ($\Sigma_{\rm SFR}$) decreases with $R_{\rm gc}$ (ref.~\citen{2016ApJ...833..229L,2022ApJ...941..162E}), so the effect of stellar feedback effect should also be weak in the outer Galaxy. 

To examine the star-formation activities in the outer-disk clouds studied in this work, we compared mid-infrared (3.4, 4.6, 12 and 22 $\mu$m wavelength) continuum images from the Wide-field Infrared Survey Explorer archive and radio continuum images from the Very Large Array Sky Survey (3 GHz S-band) and the National Radio Astronomy Observatory's Very Large Array Sky Survey (1.4 GHz L-band). Four clouds did not have archival radio data.

Although most of the outer-disk molecular clouds show obvious 22 $\mu$m emissions, only a few of them show obvious 3 and 1.4 GHz continuum emission (`Data Availability'). Therefore, the outer-disk molecular clouds studied in this work are mostly star-forming clumps\cite{lada1987star} without strong massive stellar feedback (otherwise, the 3 and 1.4 GHz continua would show strong free-free emission from \hii\ regions and the velocity dispersion would be increased by the massive stellar feedback). Because the feedback from low-mass stars does not influence the overall cloud dynamics\cite{2012ApJ...747...22H,2015ApJS..219...20L}, the $\alpha_{\rm vir}$ trends should also apply to non-star-forming clouds in general.

The subvirial outer-disk clouds at $R_{\rm gc}>15$ kpc used in this work are from ref.~\citen{2015ApJ...798L..27S}. A sensitive survey of H$_{2}$O, CH$_{3}$OH and OH masers was performed towards these clouds \cite{2018ApJ...869..148S}, but no detection was found. This is in line with the evidence that the outer-disk molecular clouds lack massive star formation and, therefore, the stellar feedback is weak.

\subsection*{Caveats of studying \alphavir\ using low-$J$ $^{12}$CO lines} 

Low-$J$ $^{12}$CO lines, with the standard \Xco\ conversion factor, have been widely adopted to trace molecular gas conditions among \Htwo\ clouds in both the Milky Way and external galaxies\cite{2013ARA&A..51..207B}. However, this method is strictly limited in large-scale studies (50 pc to kiloparsecs) \cite{2013ARA&A..51..207B}. For resolved studies of individual molecular clouds, radiative trapping becomes dominant in regulating the $^{12}$CO emission.

First, the optical depths are not uniformly distributed in the $^{12}$CO $J=1\rightarrow0$ emission (in both the spatial and velocity domains). A simple $X_{\rm CO}$ factor cannot properly trace the real distributions of mass and velocity of \Htwo\ gas. 

Second, the effective critical density of the $^{12}$CO $J=1\rightarrow0$ line is low. At a kinetic temperature of 20 K and under the optically thin limit \cite{2015PASP..127..299S}, the critical density $n_{\rm crit}$ of $^{12}$CO $J=1\rightarrow0$ is $\sim 3.5\times 10^2$\cmt\ (`Code Availability'). This value is $\times$10 lower than that often seen in the literature \cite{2013ARA&A..51..105C}, which neglects that collisions do not follow any selection rules\cite{2015PASP..127..299S}.
  
However, the $^{12}$CO $J=1\rightarrow0$ line shows a high optical depth (with a mean value $\tau > 10$, by multiplying the $^{12}$C/$^{13}$C ratio with $\tau_{\rm ^{13}CO}$) \cite{2009ApJS..182..131R} in Galactic giant molecular clouds, indicating strong local radiative trapping\cite{draine2010physics}. This radiative trapping would lower the effective critical density $n_{\rm crit}^{\rm eff}$ by another factor of $\tau$ ($>$10): $n_{\rm crit}^{\rm eff} \approx n_{\rm crit}(1-e^{-\tau})/\tau$, where $(1-e^{-\tau})/\tau \approx 1/\tau$ represents the escape probability of the transition. This means that \Htwo\ gas with a low density of a few $\times$ 10\,\cmt\ could remarkably contribute to the $^{12}$CO $J=1\rightarrow0$ emission in both line flux and linewidth.

\begin{figure}[t!]
    \centering
    \includegraphics[width=\linewidth]{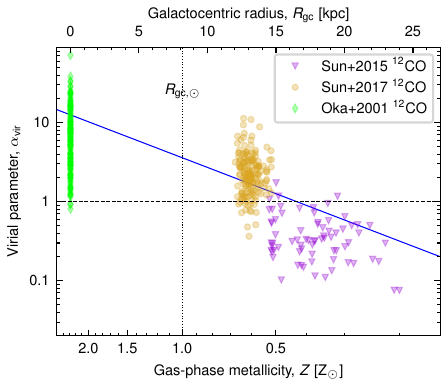}
    \caption{{\bf Variation of \alphavir\ with $Z$ and $R_{\rm gc}$.} The \alphavir\ are measured using $^{12}$CO emission from \cite{2015ApJ...798L..27S,2017ApJS..230...17S,2001ApJ...562..348O}. The blue line shows a linear fitting of the \alphavir$-Z$ trend using the \thco\ data in the Milky Way (see Methods). Data are presented as measured values with 1-$\sigma$ uncertainties.}
\label{fig: ED_Fig3}
\end{figure}

As a sanity check, we also collected $^{12}$CO measurements from the literature \cite{2015ApJ...798L..27S,2017ApJS..230...17S,2001ApJ...562..348O}. We corrected $M_{\rm mol}$ by implementing the metallicity-dependent $X_{\rm CO}$. We simply adopted $X_{\rm CO} \propto Z^{-1.0}$ as the dependency relation \cite{2013ARA&A..51..207B}. Supplementary Fig.~\ref{fig: ED_Fig3} shows that the \alphavir\ measured using $^{12}$CO emission follows a similar trend \thco: $\log(\alpha_{\rm vir}) = (1.5\pm0.1)\times\log(Z [Z_{\odot}])+(0.56\pm0.03)$. This correlation was fitted using ordinary least squares for the $^{13}$CO measurements of the Galactic molecular clouds (Fig.~\ref{fig: Main_Fig3}). The slope of the \alphavir\ versus $Z$ trend is sensitive to the adopted \Htwo/\thco\ abundance ratio, which may be better constrained by future observations.

\section*{Data Availability}

The following data and figures are available via figshare at \url{https://doi.org/10.6084/m9.figshare.27282924} (ref.\citen{Lin2024}): (1) reduced data cubes of the new observations presented in this work (subfolder GMPMC\_line\_fitscubes), (2) $^{13}$CO spectra and $N_{\rm H_2}$ maps (subfolder Supp\_Figures/NH2\_13CO\_spectrum), (3) infrared and radio images (subfolder Supp\_Figures/IR\_Radio\_outer\_disk\_clouds) and (4) distance PDFs of the Galactic molecular clouds (subfolder Supp\_Figures/Distance\_PDFs). This work is based on observations carried out under Projects 031-17 and 102-22 with the IRAM 30 m telescope, Projects ADS/JAO.ALMA\#2013.1.00652.S, ADS/JAO.ALMA\#2015.1.00581.S, ADS/JAO.ALMA\#2019.1.01641.S and ADS/JAO.ALMA\#2021.2.00175.S with ALMA, and Project Lin\_L\_22B\_1 with SMT. Source data are provided with this paper.

\section*{Code Availability}

Code for calculating the critical densities can be obtained from GitHub
(\url{https://github.com/ZhiyuZhang/critical_densities}).

\section*{Acknowledgements}

This work is supported by the National Key Research \& Development
(R\&D) Programme of China (Grant No. 2023YFA1608204). Z.-Y.Z.,
L. Lin, Yichen Sun and G.L. acknowledge the support of the National
Natural Science Foundation of China (NSFC; Grant Nos. 12173016 and
12041305), science research grants from the China Manned Space
Project (Grant Nos. CMS-CSST-2021-A08 and CMS-CSST-2021-A07)
and the Programme for Innovative Talents, Entrepreneur in Jiangsu.
This work also benefited from the International Space Science
Institute (ISSI/ISSI-BJ) in Bern and Beijing, thanks to funding for the
team ‘Chemical abundances in the ISM: the litmus test of stellar
IMF variations in galaxies across cosmic time’ (PIs D.R. and Z.-Y.Z.).
J. Wang gives thanks for the support of the NSFC (Grant No. 12173067)
and the Guangxi Talent Programme (Highland of Innovation Talents).
Y.G. is supported by the Strategic Priority Research Programme
of the Chinese Academy of Sciences (Grant No. XDB0800301).
Yan Sun is supported by the Youth Innovation Promotion Association,
CAS (Grant No. 2022085), and the Light of West China Programme
(Grant No. xbzg-zdsys-202212). T.G.B. acknowledges support from
the Leading Innovation and Entrepreneurship Team of Zhejiang
Province of China (Grant No. 2023R01008). D.R. thanks the Italian
National Institute for Astrophysics for funding the project ‘An
in-depth theoretical study of CNO element evolution in galaxies’
through Finanziamento della Ricerca Fondamentale, Theory Grant
Fu. Ob. 1.05.12.06.08. D.L. is a New Cornerstone investigator. H.B.L. is
supported by the National Science and Technology Council of Taiwan
(Grant Nos. 111-2112-M-110-022-MY3 and 113-2112-M-110-022-MY3).
K.Q. acknowledges support from the NSFC (Grant Nos. 12425304
and U1731237) and the National Key R\&D Programme of China (Grant
Nos. 2023YFA1608204 and 2022YFA1603100). C.-W.T. is supported
by the NSFC (Grant No. 11988101). J. Wu gives thanks for support from
the NSFC (Grant No. 12041302) and the Tianchi Talent Programme of
Xinjiang Uygur Autonomous Region. S.F. acknowledges support from
the NSFC (Grant No. 12373023), a starting grant at Xiamen University
and the presidential excellence fund at Xiamen University (Grant No.
20720220024). This work is based on observations carried out under
Projects 031-17 and 102-22 with the IRAM 30 m telescope, Projects
ADS/JAO.ALMA\#2013.1.00652.S, ADS/JAO.ALMA\#2015.1.00581.S, ADS/
JAO.ALMA\#2019.1.01641.S and ADS/JAO.ALMA\#2021.2.00175.S with
ALMA, and Project Lin\_L\_22B\_1 with SMT. IRAM is supported by INSU/
CNRS (France), MPG (Germany) and IGN (Spain). ALMA is a partnership
of ESO (representing its member states), NSF (USA) and NINS (Japan),
together with NRC (Canada), MOST and ASIAA (Taiwan) and KASI
(Republic of Korea), in cooperation with the Republic of Chile. The
Joint ALMA Observatory is operated by ESO, AUI/NRAO and NAOJ.
The Heinrich Hertz SMT is operated by the Arizona Radio Observatory,
which is part of Steward Observatory at the University of Arizona.

\section*{Author Contributions statement}

L. Lin led the project, conducted the data reduction and analyses, and
drafted proposals and the paper. Z.-Y.Z. initiated and supervised the
whole project and improved the paper. J. Wang helped with the data
analysis and validating tests. P.P.P. instructed the general virial analysis
and outlined the larger theoretical picture. Yong Shi helped extend this
work to low-metallicity dwarf galaxies. Y.G. helped improve the paper
and was involved in discussions. Yan Sun provided the initial catalogue
and helped with calculating the cloud distances. Yichen Sun, T.G.B.
and D.R. helped with the abundance ratios. D.L., H.B.L. and K.Q. helped
with discussions on magnetic fields. B.Z. helped with validating the
distance measurements. L. Liu, G.L., C.-W.T., J. Wu and S.F. helped
with discussions and in improving the paper. All authors reviewed
the paper and were involved in discussions, telescope proposals and
observations on which the raw data and the analyses were based.

\section*{Competing Interests Statement}
The authors declare no competing interests.


\clearpage

%
%

\begin{table*}[h!]
\centering
\caption{Measured quantities for molecular clouds observed by IRAM 30-m and SMT. *: Clouds with strong \textit{VLASS} or \textit{NVSS} continuum.}
\label{tab: Supp_Table1}
\resizebox{\linewidth}{!}{
\begin{tabular}{ccccccccccc}
\hline
Target       & $l$          & $b$          & $V_{\rm LSR}$ & $R_{\rm gc}$ & $d$   & $R_{\rm cloud}$ & $\sigma_{\rm v}$ & $M_{\rm cloud}$        & $n_{\rm H_{2}}$      & $\alpha_{\rm vir}$ \\
             & ($^{\circ}$) & ($^{\circ}$) & (km s$^{-1}$) & (kpc)        & (kpc) & (pc)            & (km s$^{-1}$)    & (10$^{3}$ $M_{\odot}$) & (10$^{4}$ cm$^{-3}$) & \\
\hline
\multicolumn{11}{c}{IRAM 30-m (2016)} \\
\hline
G37.350    & 37.35  & 1.06  & -54.0  & 13.2$\pm$1   & 18.7$\pm$1  & 1.0$\pm$0.1  & 0.9$\pm$0.09 & 1.1$\pm$0.2  & 0.3 & 1.0$\pm$0.3 \\
G44.8      & 44.80  & 0.66  & -61.9  & 13.0$\pm$1   & 17.5$\pm$1  & 0.9$\pm$0.1  & 1.0$\pm$0.1  & 0.9$\pm$0.2  & 0.5 & 1.1$\pm$0.3 \\
IRAS0245   & 136.35 & 0.96  & -61.5  & 13.2$\pm$1   & 6.0$\pm$1   & 0.4$\pm$0.08 & 0.7$\pm$0.07 & 0.4$\pm$0.07 & 2   & 0.7$\pm$0.3 \\
SUN15\_14N & 109.29 & 2.08  & -101.1 & 14.9$\pm$1   & 10.1$\pm$1  & 0.9$\pm$0.1  & 0.5$\pm$0.05 & 0.4$\pm$0.08 & 0.2 & 0.7$\pm$0.2 \\
SUN15\_18  & 109.79 & 2.71  & -99.1  & 14.7$\pm$0.9 & 9.8$\pm$1   & 0.7$\pm$0.1  & 0.4$\pm$0.04 & 0.3$\pm$0.07 & 0.4 & 0.4$\pm$0.1 \\
SUN15\_21* & 114.34 & 0.79  & -101.0 & 15.4$\pm$1   & 10.1$\pm$1  & 1.4$\pm$0.2  & 0.5$\pm$0.05 & 0.8$\pm$0.2  & 0.1 & 0.6$\pm$0.2 \\
SUN15\_34  & 122.77 & 2.52  & -107.0 & 17.8$\pm$2   & 12.0$\pm$2  & 1.1$\pm$0.2  & 0.5$\pm$0.05 & 0.8$\pm$0.2  & 0.2 & 0.4$\pm$0.1 \\
SUN15\_56  & 137.77 & -0.97 & -103.1 & 24.1$\pm$4   & 17.4$\pm$4  & 2.7$\pm$0.6  & 0.7$\pm$0.07 & 13.6$\pm$3   & 0.3 & 0.1$\pm$0.04 \\
SUN15\_57  & 137.78 & -1.07 & -102.2 & 23.7$\pm$3   & 17.0$\pm$4  & 1.4$\pm$0.3  & 0.6$\pm$0.06 & 3.2$\pm$0.6  & 0.4 & 0.2$\pm$0.07 \\
SUN15\_7W  & 104.98 & 3.31  & -102.6 & 14.8$\pm$0.9 & 10.5$\pm$1  & 0.7$\pm$0.1  & 0.6$\pm$0.06 & 0.3$\pm$0.07 & 0.4 & 0.9$\pm$0.3 \\
WB89\_380* & 124.65 & 2.54  & -86.2  & 14.7$\pm$1   & 8.5$\pm$1   & 1.1$\pm$0.2  & 1.4$\pm$0.1  & 2.2$\pm$0.4  & 0.5 & 1.2$\pm$0.4 \\
WB89\_391* & 125.80 & 3.05  & -86.1  & 14.9$\pm$1   & 8.6$\pm$1   & 0.5$\pm$0.09 & 0.7$\pm$0.07 & 0.4$\pm$0.08 & 1   & 0.7$\pm$0.2 \\
WB89\_437  & 135.28 & 2.80  & -71.8  & 13.4$\pm$0.6 & 6.3$\pm$0.6 & 0.5$\pm$0.07 & 1.2$\pm$0.1  & 0.5$\pm$0.1  & 1   & 1.5$\pm$0.5 \\
WB89\_501  & 145.20 & 2.98  & -58.2  & 14.6$\pm$1   & 7.1$\pm$2   & 0.3$\pm$0.07 & 0.8$\pm$0.08 & 0.3$\pm$0.07 & 4   & 0.7$\pm$0.3 \\
\hline
\multicolumn{11}{c}{IRAM 30-m (2023)} \\
\hline
SUN15\_23 & 116.72 & 3.54  & -107.2 & 16.6$\pm$1 & 11.3$\pm$1 & 0.7$\pm$0.1  & 0.6$\pm$0.06 & 0.4$\pm$0.07 & 0.4  & 0.8$\pm$0.3 \\
SUN15\_25 & 117.58 & 3.95  & -106.0 & 16.6$\pm$1 & 11.2$\pm$1 & 0.3$\pm$0.05 & 0.5$\pm$0.05 & 0.2$\pm$0.04 & 2    & 0.5$\pm$0.2 \\
SUN15\_30 & 121.81 & 3.05  & -104.1 & 17.0$\pm$1 & 11.3$\pm$2 & 0.8$\pm$0.1  & 0.6$\pm$0.06 & 0.5$\pm$0.1  & 0.3  & 0.7$\pm$0.2 \\
SUN15\_53 & 137.29 & -1.16 & -101.5 & 23.0$\pm$3 & 16.3$\pm$3 & 2.3$\pm$0.5  & 0.5$\pm$0.05 & 1.4$\pm$0.3  & 0.04 & 0.4$\pm$0.2 \\
SUN15\_55 & 137.62 & -1.23 & -98.8  & 22.1$\pm$3 & 15.4$\pm$3 & 1.1$\pm$0.2  & 0.5$\pm$0.05 & 0.9$\pm$0.2  & 0.2  & 0.4$\pm$0.1 \\
SUN15\_59 & 139.12 & -1.47 & -96.6  & 22.4$\pm$3 & 15.6$\pm$3 & 0.5$\pm$0.1  & 0.4$\pm$0.04 & 0.5$\pm$0.1  & 1    & 0.2$\pm$0.06 \\
SUN15\_69 & 145.21 & -0.39 & -74.0  & 18.5$\pm$2 & 11.2$\pm$2 & 1.2$\pm$0.3  & 0.5$\pm$0.05 & 0.4$\pm$0.08 & 0.08 & 0.8$\pm$0.3 \\
SUN15\_72 & 146.06 & -1.65 & -77.4  & 20.3$\pm$3 & 13.1$\pm$3 & 1.9$\pm$0.5  & 0.5$\pm$0.05 & 0.9$\pm$0.2  & 0.05 & 0.6$\pm$0.2 \\
\hline
\multicolumn{11}{c}{SMT (2022)} \\
\hline
SUN15\_47 & 131.16 & 1.39 & -100.4 & 19.0$\pm$2 & 12.6$\pm$2 & 0.2$\pm$0.03 & 0.3$\pm$0.03 & 0.1$\pm$0.03 & 8   & 0.2$\pm$0.06 \\
WB89\_379 & 124.56 & 2.52 & -88.1  & 15.0$\pm$1 & 8.8$\pm$1  & 0.5$\pm$0.09 & 0.9$\pm$0.09 & 0.3$\pm$0.07 & 0.9 & 1.5$\pm$0.5 \\
WB89\_434 & 135.99 & 0.67 & -78.0  & 15.7$\pm$1 & 8.8$\pm$2  & 0.5$\pm$0.09 & 0.5$\pm$0.05 & 0.1$\pm$0.01 & 0.3 & 2.0$\pm$0.7 \\
WB89\_365 & 123.07 & 1.35 & -90.9  & 15.1$\pm$1 & 9.1$\pm$1  & 0.5$\pm$0.08 & 0.5$\pm$0.05 & 0.1$\pm$0.01 & 0.2 & 2.6$\pm$0.8 \\
\hline
\end{tabular}
}
\end{table*}

\begin{table*}[h!]
\centering
\caption{Measured quantities for molecular clouds observed by ALMA Total Power. *: Clouds with strong \textit{VLASS} or \textit{NVSS} continuum. $\dag$: Clouds outside of the sky coverage of VLA.} 
\label{tab: Supp_Table2}
\resizebox{\linewidth}{!}{
\begin{tabular}{ccccccccccc}
\hline
Target        & $l$          & $b$          & $V_{\rm LSR}$ & $R_{\rm gc}$ & $d$   & $R_{\rm cloud}$ & $\sigma_{\rm v}$ & $M_{\rm cloud}$        & $n_{\rm H_{2}}$      & $\alpha_{\rm vir}$ \\
              & ($^{\circ}$) & ($^{\circ}$) & (km s$^{-1}$) & (kpc)        & (kpc) & (pc)            & (km s$^{-1}$)    & (10$^{3}$ $M_{\odot}$) & (10$^{4}$ cm$^{-3}$) & \\
\hline
\multicolumn{11}{c}{ ALMA Total Power (Cycle 8)} \\
\hline
WB89\_766        & 193.43 & -1.17 & 22.3 & 14.4$\pm$3   & 6.4$\pm$3   & 0.6$\pm$0.3  & 1.1$\pm$0.1  & 0.4$\pm$0.07 & 0.5 & 2.3$\pm$1 \\
WB89\_770*       & 192.91 & -0.63 & 22.6 & 15.0$\pm$3   & 7.0$\pm$3   & 0.4$\pm$0.2  & 0.9$\pm$0.09 & 0.3$\pm$0.07 & 3   & 1.0$\pm$0.5 \\
WB89\_1086       & 245.93 & 1.17  & 63.5 & 11.6$\pm$0.6 & 5.6$\pm$0.8 & 0.8$\pm$0.1  & 0.7$\pm$0.07 & 0.3$\pm$0.06 & 0.2 & 1.8$\pm$0.6 \\
WB89\_765*       & 196.83 & -3.11 & 25.7 & 12.2$\pm$0.2 & 4.1$\pm$0.2 & 0.4$\pm$0.05 & 0.8$\pm$0.08 & 0.2$\pm$0.05 & 1   & 1.5$\pm$0.5 \\
WB89\_771        & 193.69 & -1.06 & 22.9 & 14.5$\pm$3   & 6.4$\pm$3   & 0.7$\pm$0.3  & 0.9$\pm$0.09 & 0.2$\pm$0.04 & 0.3 & 2.7$\pm$1 \\
WB89\_1010       & 230.55 & 1.92  & 63.1 & 11.7$\pm$0.3 & 4.7$\pm$0.3 & 0.4$\pm$0.04 & 0.6$\pm$0.06 & 0.1$\pm$0.02 & 0.5 & 2.1$\pm$0.6 \\
WB89\_1049*      & 241.54 & -0.60 & 70.5 & 11.9$\pm$0.4 & 5.6$\pm$0.5 & 0.9$\pm$0.1  & 0.8$\pm$0.08 & 0.3$\pm$0.07 & 0.2 & 1.9$\pm$0.6 \\
WB89\_1048*      & 240.32 & 0.07  & 67.8 & 11.9$\pm$0.4 & 5.5$\pm$0.5 & 0.4$\pm$0.06 & 2.4$\pm$0.2  & 0.4$\pm$0.09 & 2   & 6.5$\pm$2 \\
WB89\_1158$\dag$ & 260.69 & -1.39 & 66.3 & 11.5$\pm$0.5 & 6.8$\pm$0.8 & 1.0$\pm$0.2  & 1.2$\pm$0.1  & 0.4$\pm$0.09 & 0.1 & 3.8$\pm$1 \\
WB89\_782*       & 196.46 & -1.68 & 17.1 & 12.2$\pm$0.2 & 4.1$\pm$0.2 & 0.6$\pm$0.07 & 1.5$\pm$0.2  & 0.7$\pm$0.1  & 1   & 2.3$\pm$0.7 \\
WB89\_959        & 231.52 & -4.30 & 52.9 & 11.5$\pm$0.7 & 4.5$\pm$0.8 & 0.6$\pm$0.1  & 1.0$\pm$0.1  & 0.2$\pm$0.04 & 0.3 & 3.5$\pm$1 \\
WB89\_886        & 212.06 & -0.74 & 44.6 & 11.9$\pm$0.2 & 4.2$\pm$0.2 & 0.3$\pm$0.04 & 1.6$\pm$0.2  & 0.1$\pm$0.03 & 1   & 7.4$\pm$2 \\
WB89\_905        & 211.59 & 1.06  & 44.8 & 11.9$\pm$0.2 & 4.2$\pm$0.2 & 0.4$\pm$0.04 & 1.6$\pm$0.2  & 0.2$\pm$0.04 & 1   & 5.7$\pm$2 \\
WB89\_1236$\dag$ & 269.22 & -0.31 & 74.8 & 12.0$\pm$0.6 & 8.7$\pm$0.8 & 0.8$\pm$0.1  & 0.7$\pm$0.07 & 0.2$\pm$0.04 & 0.1 & 2.2$\pm$0.7 \\
WB89\_762        & 190.97 & -0.06 & 11.0 & 11.1$\pm$3   & 3.0$\pm$3   & 0.5$\pm$0.5  & 0.5$\pm$0.05 & 0.1$\pm$0.01 & 0.2 & 2.1$\pm$2 \\
WB89\_904        & 211.04 & 1.18  & 55.2 & 15.0$\pm$2   & 7.4$\pm$2   & 1.1$\pm$0.3  & 0.8$\pm$0.08 & 0.6$\pm$0.1  & 0.1 & 1.6$\pm$0.6 \\
WB89\_864        & 210.24 & -1.33 & 37.0 & 12.0$\pm$1   & 4.2$\pm$1   & 0.3$\pm$0.09 & 0.8$\pm$0.08 & 0.0$\pm$0.01 & 0.6 & 4.7$\pm$2 \\
WB89\_985N       & 229.57 & 0.15  & 53.2 & 11.6$\pm$0.2 & 4.6$\pm$0.3 & 0.5$\pm$0.06 & 1.4$\pm$0.1  & 0.2$\pm$0.05 & 0.6 & 4.9$\pm$2 \\
WB89\_985S       & 229.57 & 0.15  & 53.2 & 11.6$\pm$0.2 & 4.6$\pm$0.3 & 0.5$\pm$0.06 & 1.0$\pm$0.1  & 0.2$\pm$0.04 & 0.5 & 3.0$\pm$0.9 \\
WB89\_885        & 212.35 & -0.92 & 44.3 & 11.9$\pm$0.2 & 4.2$\pm$0.2 & 0.8$\pm$0.09 & 1.2$\pm$0.1  & 0.3$\pm$0.06 & 0.2 & 4.3$\pm$1 \\
WB89\_1145$\dag$ & 259.24 & -1.63 & 62.7 & 11.2$\pm$0.5 & 6.4$\pm$0.8 & 0.3$\pm$0.05 & 0.7$\pm$0.07 & 0.0$\pm$0.01 & 0.5 & 4.1$\pm$1 \\
WB89\_1245$\dag$ & 269.85 & -0.06 & 74.4 & 12.0$\pm$0.6 & 8.7$\pm$0.8 & 0.9$\pm$0.1  & 1.6$\pm$0.2  & 0.7$\pm$0.1  & 0.3 & 4.0$\pm$1 \\
WB89\_920        & 215.19 & 0.81  & 48.7 & 12.8$\pm$1   & 5.2$\pm$1   & 0.9$\pm$0.2  & 0.8$\pm$0.08 & 0.4$\pm$0.08 & 0.2 & 1.8$\pm$0.7 \\
WB89\_1057       & 242.94 & -0.45 & 63.6 & 11.7$\pm$0.6 & 5.5$\pm$0.8 & 0.6$\pm$0.1  & 1.6$\pm$0.2  & 0.2$\pm$0.04 & 0.3 & 8.8$\pm$3 \\
WB89\_758        & 193.31 & -1.45 & 28.7 & 18.5$\pm$4   & 10.5$\pm$4  & 0.7$\pm$0.3  & 0.7$\pm$0.07 & 0.3$\pm$0.07 & 0.3 & 1.2$\pm$0.6 \\
WB89\_916        & 212.96 & 1.30  & 42.4 & 11.9$\pm$0.2 & 4.2$\pm$0.2 & 0.8$\pm$0.09 & 0.9$\pm$0.09 & 0.2$\pm$0.05 & 0.2 & 2.9$\pm$0.9 \\
\hline
\end{tabular}
}
\end{table*}

\clearpage

%
%
%

\end{document}